\documentclass[twocolumn]{aastex701}

\usepackage{amsmath}
\usepackage{multirow}
\usepackage{booktabs}

\newcommand{\simba}{\textsc{Simba}}
\newcommand{\tng}{\textsc{IllustrisTNG}}
\newcommand{\eagle}{\textsc{Eagle}}
\newcommand{\eaglesm}{\textsc{Eagle}\,RefL0025}
\newcommand{\code}[1]{\texttt{#1}}
\newcommand{\msun}{\text{M}_{\odot}}
\newcommand{\lsun}{\text{L}_{\odot}}

\renewcommand{\tablecomments}[1]{%
  \par\vspace{0.5ex}%
  \noindent\parbox{\linewidth}{\small\textit{Note.}~#1}%
}

\begin{document}

\title{Predicting Supermassive Black Hole-Host Mass Offsets from Broadband Photometry Across Cosmological Simulations with Forecasts for LSST}

\author[0009-0001-2531-5987]{Hitaishi Chillara}
\email{hitaishi@hitu.dev}
\affiliation{Independent Researcher}

\begin{abstract}
We forecast what broadband Legacy Survey of Space and Time (LSST) photometry encodes about the offset $\Delta$ of supermassive black hole mass from the per-simulation, per-redshift running-median $M_{\rm BH}$--$M_*$ relation. We forward-model the \simba, \tng, and \eagle\ simulations into LSST $ugrizy$ photometry and evaluate nested regressors cross-validated by black hole identifier. Broadband colors contain a resolved fraction of $\Delta$: for \simba\ the full photometry reduces the residual scatter from $0.598$ to $0.505$ dex in the magnitude-limited sample, a $15.7\%$ reduction ($R^2 = 0.277$), and from $0.697$ to $0.653$ dex at the intrinsic level, a $6.2\%$ reduction ($R^2 = 0.103$). A regressor given the true stellar mass and redshift recovers nothing ($R^2 = -0.014$). The recovery is positive in every volume, with intrinsic scatter reductions of $4.8\%$ to $17.4\%$, and the color increment keeps its sign and order of magnitude across a factor-of-four range in assumed host scatter. The signal splits into a host stellar-population channel and an AGN-brightness channel; the color-driven constraint strengthens $4.8$-fold from $z < 0.5$ to $z > 5$. The stellar-population channel dominates \simba's intrinsic signal and the AGN channel leads in \tng, an ordering consistent with, though not fixed by, the black hole mass dependence of their accretion prescriptions. Cross-simulation transfer is partial and asymmetric, degrading by $0.022$ to $0.027$ dex into \simba\ but by $0.230$ to $0.350$ dex out of it; out-of-\simba\ predictions can be anti-informative. The recovery is resolved but small; per-object $\Delta$ at the precision of local scaling relations requires spectroscopic follow-up.
\end{abstract}


\section{Introduction} \label{sec:intro}

The Vera C. Rubin Observatory's Legacy Survey of Space and Time (LSST) will image billions of galaxies in $ugrizy$ broadband photometry, nearly all of which will lack direct black hole mass measurements \citep{Ivezic2019}. What this imaging can constrain about the residual offset between a galaxy's central supermassive black hole (SMBH) mass and the population-level $M_{\rm BH}$--$M_*$ relation is an observational question.

The history of accretion and black hole-host co-evolution leaves persistent imprints on the $M_{\rm BH}$--$M_*$ relation \citep{Habouzit2021,Habouzit2022,Pacucci2023,Sassano2021,Bhowmick2025}. JWST has revealed populations of luminous high-redshift AGN \citep{Harikane2023,Maiolino2024,Kocevski2023} that lie above local scaling relations, and the Trinity empirical model \citep{Zhang2023Trinity,Zhang2024TrinityIV} finds that luminosity-selected samples at high redshift are biased toward systematically over-massive systems by $0.35$--$1.0$ dex relative to the population median. Locally, the $M_{\rm BH}$--$M_*$ relation is calibrated by dynamical and reverberation black hole masses across the galaxy population \citep{ReinesVolonteri2015,Greene2020}. Cosmological hydrodynamic simulations such as \simba\ \citep{Dave2019}, \tng\ \citep{Springel2017,Pillepich2018,Nelson2019}, and \eagle\ \citep{Schaye2015,Crain2015} utilize distinct subgrid prescriptions for SMBH seeding, accretion, and feedback, and the residual $\Delta = \log_{10} M_{\rm BH} - \log_{10} M_{\rm BH,med}(M_*, z)$ carries information about the local growth channel \citep{Inayoshi2020,Volonteri2021,Pacucci2024}. \citet{Piotrowska2022} train machine-learning classifiers to predict galaxy quiescence across \simba, \tng, and \eagle\ and find central black hole mass to be the dominant predictor, motivating the same suite to ask how much residual $M_{\rm BH}$ information survives transmission into broadband photometry.

We forecast how well broadband LSST photometry can recover $\Delta$ at the level of individual galaxies, forward-modeling \simba, \tng, and \eagle\ into LSST photometry, defining $\Delta$ against per-simulation per-redshift running-median reference relations to avoid the curvature and redshift evolution embedded in a single linear fit, and regressing on continuous $\Delta$ with an ensemble of four base models stacked through a meta-learner. The methodological contribution is fourfold. First, a nested set of regressors whose key member is given the true stellar mass and redshift is an orthogonalization standard: a regressor with exact $M_*$ and $z$ that recovers nothing shows the signal is not recovered stellar mass. Second, the same regressors' host-only and accretion-ablation machinery decomposes the photometric information about $\Delta$ into a stellar-population channel and an AGN-light channel. Third, the AGN channel's informativeness differs across simulations, from consistent with zero in the small \eagle\ volume to dominant in \tng, an ordering consistent with the nominal black hole mass dependence of each accretion prescription but not fixed by it, since the tabulated accretion rates also depend on nuclear gas conditions the catalogs do not resolve. Fourth, we produce the LSST-era forecast numbers and their cross-simulation transfer. The coupling's sign is preserved across simulations while its amplitude is simulation-specific.

\section{Data} \label{sec:data}

\subsection{Simulations and Data Selection}

We draw upon data from the \simba\ \citep{Dave2019} and \tng\ \citep{Pillepich2018,Nelson2019,Springel2017,Marinacci2018,Naiman2018,Nelson2019DataRelease} cosmological simulations. \simba\ evolved a $(100~h^{-1}~{\rm cMpc})^3$ comoving volume from $z = 249$ to $z = 0$ using the \textsc{Gizmo} code \citep{Hopkins2015}, seeding SMBHs at $10^4~\msun/h$ and using a two-mode accretion model \citep{Hopkins2011,AnglesAlcazar2017}. TNG100-1 evolved a $(75~h^{-1}~{\rm cMpc})^3$ volume using the \textsc{AREPO} code \citep{Springel2010}, seeding SMBHs at $\sim 1.2 \times 10^6~\msun$ with Eddington-limited Bondi-Hoyle accretion and dual-mode AGN feedback \citep{Weinberger2017,Weinberger2018}. We also analyze TNG300-1 for resolution and volume tradeoffs.

From \simba, we obtain BH catalogs that contain $\sim 20$ million entries spanning $z = 0$ to $z \approx 10$, of which $\sim 2$ million lie at $z \geq 4$. All entries include: black hole mass ($M_{\rm BH}$), mass accretion rate ($\dot{M}$), host galaxy stellar mass ($M_*$), host gas mass ($M_{\rm gas}$), star formation rate (SFR), scale factor ($a$), and a unique identifier.

Each \simba\ entry is one black hole at one snapshot. The catalog retains every black hole: a galaxy hosting more than one contributes one entry per black hole, each carrying the same host $M_*$, $M_{\rm gas}$, and SFR, and we apply no selection of the most massive or most central black hole per galaxy. Such multiple-occupancy hosts are the source of the shared-host contamination across cross-validation folds, which is below $0.3\%$ (Section~\ref{sec:features}), so their weight in the running-median reference is correspondingly negligible.

From the \tng\ simulation suite, we collect entries that span $z = 0$ to $z = 10$, including 17,254 entries from TNG100-1 and 31,762 from TNG300-1. Each \tng\ entry is a SUBFIND subhalo rather than a black hole particle: its black hole mass and accretion rate are the subhalo totals over the black hole particles it contains (\code{SubhaloBHMass}, \code{SubhaloBHMdot}), so a galaxy with several black hole particles enters once, with a mass dominated by its most massive black hole. Catalog validity cuts ($M_{\rm BH} > 10^4~\msun$, $M_* > 10^6~\msun$), a forward-model accretion floor ($\dot{M} > 10^{-10}~\msun\,{\rm yr}^{-1}$), and the requirement of finite forward-modeled photometry reduce these to the 15,242 and 28,074 intrinsic samples of Table~\ref{tab:comparison}.

The accretion floor is a requirement of the forward model, not a physical validity criterion: black holes below it are valid catalog entries, in gas-poor hosts where the simulation no longer resolves the tenuous nuclear gas. The AGN component of Section~\ref{sec:methods} takes logarithms of $L_{\rm bol,AGN}$ and $\lambda_{\rm Edd}$ in its ultraviolet-slope and Eddington-ratio terms, which are undefined at $\dot{M} = 0$, and the floor is the guard for those terms. At the floor $L_{\rm bol,AGN} \approx 6 \times 10^{35}~{\rm erg\,s^{-1}}$ ($\sim 150~\lsun$), more than two orders of magnitude below the light of the faintest host admitted by the stellar-mass cut in every LSST band, so the excluded objects would have entered the photometry as pure hosts. Their exclusion truncates the quiescent end of the host distribution, which is where the host color--$\Delta$ coupling of Section~\ref{sec:circularity} places over-massive black holes; the floor therefore removes hosts that carry the stellar-population signal rather than adding to it, and we have not quantified its effect on the increments. In the \simba\ magnitude-limited sample the $\lambda_{\rm Edd} > 0.01$ floor (Section~\ref{sec:features}) supersedes it by more than four orders of magnitude in $\dot{M}$ at every black hole mass in the catalog.

Both simulation suites are analyzed across their full redshift range; high-redshift subsets ($z \geq 4$) are examined separately where noted (Section~\ref{sec:results}). The lower-resolution TNG100-2 and TNG300-2 variants are drawn from snapshot-stacked catalogs of 2,566,512 and 10,097,043 entries; after the same cuts and floor, seeded draws of 250,000 rows supply the intrinsic samples of the resolution study of Appendix~\ref{sec:appendix_resolution}, while its magnitude-limited samples retain every row passing the survey cuts.

\eagle\ \citep{Schaye2015,Crain2015} provides a third independent point of comparison. Its flagship RefL0100 volume covers a $(100~{\rm Mpc})^3$ comoving box, and the smaller RefL0025 volume covers a $(25~{\rm Mpc})^3$ box at the same numerical resolution; SMBHs in both runs are seeded with a heavy initial mass of $1.475 \times 10^5~\msun$ in halos exceeding a fixed friends-of-friends mass threshold, accretion follows a Bondi-Hoyle prescription with an angular-momentum correction \citep{RosasGuevara2015}, and feedback is implemented as stochastic thermal energy injection into the surrounding gas. As in \tng, each \eagle\ entry is a subhalo carrying the database totals \code{BlackHoleMass} and \code{BlackHoleMassAccretionRate} over its black hole particles. The high seed mass yields a distinct $\Delta$ distribution at fixed $M_*$, but per-simulation reference relations (Section~\ref{sec:ml_framework}) absorb this into the within-simulation residual without breaking the regression target. Quantitative regression results in this paper use both the RefL0025 and the flagship RefL0100 volumes (Appendix~\ref{sec:appendix_eagle}).

Throughout the figures and quantitative results that follow, \simba\ is the representative simulation; \tng\ and \eagle\ are used primarily for the cross-simulation transfer experiments (Section~\ref{sec:cross_sim_results}).

\subsection{LSST DP1 Observational Baseline}

\begin{figure*}[tbp!]
\plotone{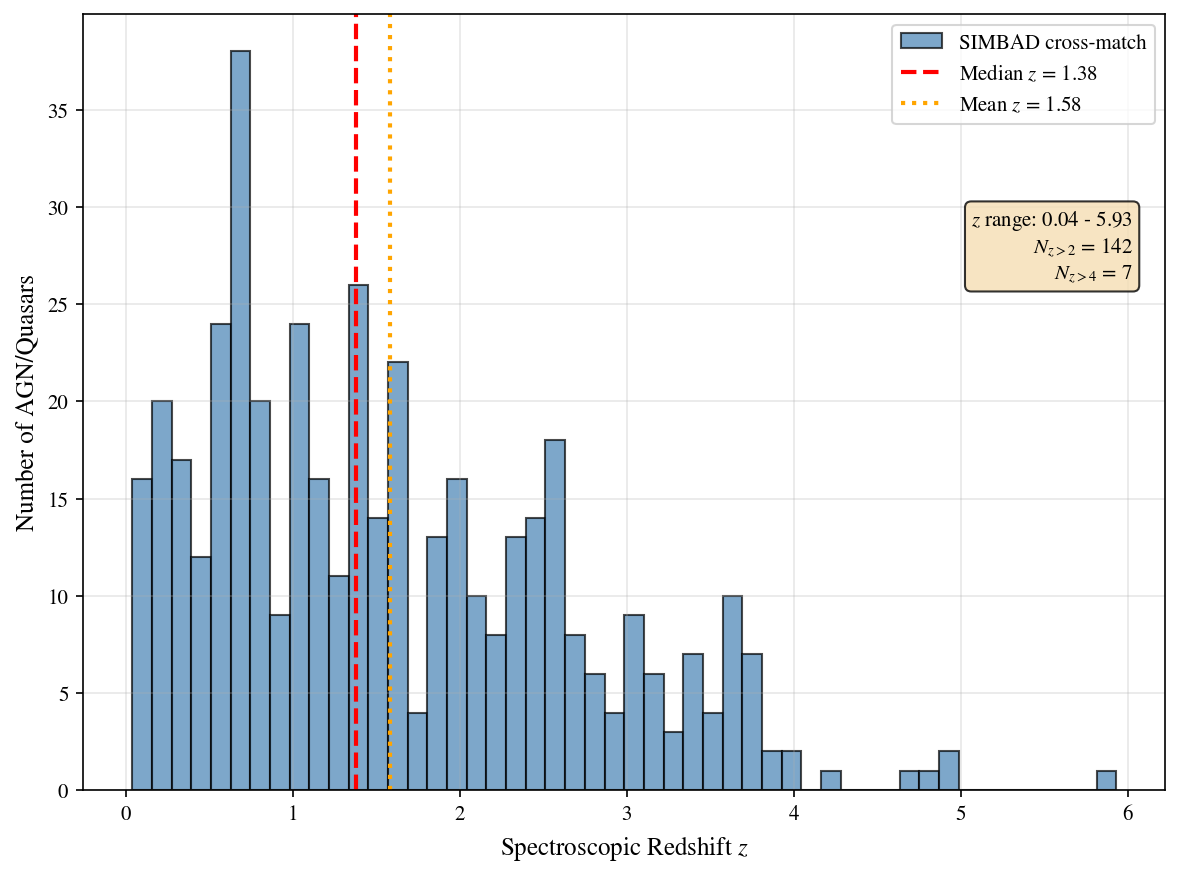}
\caption{Redshift distribution of LSST DP1 AGN candidates ($N = 439$, median $z = 1.38$). These are spectroscopic, heterogeneous literature redshifts gathered through the SIMBAD cross-match; they are not a DP1 photometric-redshift product.}
\label{fig:dp1_zdist}
\end{figure*}

For establishing an observational baseline for our forward-modeled forecasts, we construct a comparison sample from LSST Data Preview 1 \citep[DP1;][]{RubinDP1} observations of the Extended Chandra Deep Field South (ECDFS). Selecting extended sources that have signal-to-noise $> 5$ in the $r$-band yields 2,533 low-$z$ AGN candidates ($g - r < 1.0$), 286 high-$z$ $g$-dropout candidates, and 60 $r$-dropout candidates. Further cross-matching with SIMBAD recovers 439 unique objects with spectroscopic redshifts drawn from heterogeneous literature sources ($z = 0.04$ to $5.93$, median $z = 1.38$; Figure~\ref{fig:dp1_zdist}), of which the 58 in the low-$z$ candidate class at $z < 2$ enter the band-by-band comparison of Figure~\ref{fig:dp1_validation}.

\section{Methods} \label{sec:methods}

\subsection{Forward Modeling Pipeline}

The forward model transforms simulated physical properties into predicted LSST $ugrizy$ fluxes. We compute bolometric luminosities for AGN as $L_{\rm bol,AGN} = \eta \dot{M} c^2$ ($\eta = 0.1$). The AGN continuum is a piecewise power law motivated by the \citet{VandenBerk2001} composite quasar spectrum and the \citet{Richards2006} templates, with a spectral index $\alpha$ ($L_\nu \propto \nu^\alpha$) assigned per rest-frame wavelength:
\begin{equation} \label{eq:agn_sed}
\alpha(\lambda_{\rm rest}) = \begin{cases}
-1.8 & \lambda_{\rm rest} < 1000~\text{\AA} \\
-0.7 & 1000~\text{\AA} \le \lambda_{\rm rest} < 1400~\text{\AA} \\
-0.44 & 1400~\text{\AA} \le \lambda_{\rm rest} < 3000~\text{\AA} \\
-0.30 & 3000~\text{\AA} \le \lambda_{\rm rest} < 5000~\text{\AA} \\
-0.80 & 5000~\text{\AA} \le \lambda_{\rm rest} < 10000~\text{\AA} \\
-1.5 & \lambda_{\rm rest} \ge 10000~\text{\AA}.
\end{cases}
\end{equation}
Each band flux is evaluated from its rest-frame slope as a logarithmic offset from a single $2500$~\AA\ monochromatic reference, $\Delta m = 2.5\,\alpha(\lambda_{\rm rest})\,\log_{10}(\lambda_{\rm rest}/2500~\text{\AA})$; the segments set local slopes and do not enforce flux continuity across the breaks.
The $2500$~\AA\ monochromatic luminosity is set from $L_{\rm bol,AGN}$ with a constant bolometric correction of $5.15$, the $3000$~\AA\ correction of the \citet{Richards2006} composite SED applied here at $2500$~\AA; for the adopted $1400$--$3000$~\AA\ slope the corrections at the two wavelengths differ by $0.11$ mag. Motivated by \citet{Steffen2006}, the three ultraviolet segments receive an additive slope term $-0.1\,\log_{10}(L_{\rm bol,AGN}/10^{45}~{\rm erg\,s^{-1}})$ so that more luminous AGN are bluer in the ultraviolet, and the reference luminosity receives a further Eddington-ratio factor $1 + 0.15\,\log_{10}\lambda_{\rm Edd}$ clipped to $[0.5, 1.5]$, motivated by \citet{Lusso2012}.

For the IGM, we apply a simplified \citet{Madau1995}-style attenuation that retains the dominant Lyman-$\alpha$ term, with $\tau_\alpha = 0.0036\,(1+z)^{3.46}(\lambda_{\rm obs}/\lambda_{\alpha,{\rm obs}})^{3.46}$ blueward of redshifted Lyman-$\alpha$, increased by $1.5$ blueward of Lyman-$\beta$ and saturated below the Lyman limit; this reproduces the $g$-dropout signature of $z > 4$ AGN. We assign a luminosity-independent $50\%$ of the AGN as obscured ``Type 2'' with dust reddening from an exponential distribution (scale parameter $E(B-V) = 0.3$), attenuated with the SMC extinction curve \citep{Gordon2003}. The bolometric luminosity is not capped in the super-Eddington regime.

Host galaxy magnitudes in all six LSST bands are computed from single-burst stellar population synthesis spectra generated with the Flexible Stellar Population Synthesis code \citep{conroy09,conroy10} adopting a \citet{Chabrier2003} IMF. Each host receives a $\tau$-model star formation history with $\tau = M_*/{\rm SFR}$ clipped to $[0.1, 15]$ Gyr, from which a light-weighted age is derived; a gas-phase metallicity is assigned from the \citet{tremonti04} mass-metallicity relation at $z = 0$ with a linear shift in $\log Z$ following \citet{maiolino08}, then converted to $\log_{10}(Z/Z_\odot)$. The gas-phase oxygen abundance is a proxy for the stellar metallicity that sets the SSP colors; the two track each other along the stellar mass-metallicity relation \citep{Gallazzi2005} but are not identical. From the FSPS SSP spectra we precompute a $(\log t, \log Z/Z_\odot, \lambda_{\rm rest})$ grid of monochromatic mass-to-light ratios and interpolate it trilinearly per source at each band's rest wavelength; the implied K-correction is wavelength-resolved at the host redshift. Per-band intrinsic Gaussian scatter is added with $(\sigma_u, \sigma_g, \sigma_r, \sigma_i, \sigma_z, \sigma_y) = (0.18, 0.13, 0.10, 0.08, 0.07, 0.07)$ mag, capturing variation in stellar populations not resolved by the input parameters. The total observed flux is the sum of AGN and host galaxy components. We apply Galactic extinction \citep{Schlafly2011} with $E(B-V)_{\rm Gal} = 0.01$ and internal host attenuation following a \citet{Calzetti2000} law with $E(B-V)_{\rm host} = 0.44\,\log_{10}(1 + {\rm SFR}/10) + 0.15\,f_{\rm gas}$, clipped to $[0, 1]$, where $f_{\rm gas} = M_{\rm gas}/(M_{\rm gas} + M_*)$ is the gas mass fraction formed from the catalog gas and stellar masses; the gas-fraction term raises the dust column in gas-rich hosts at fixed SFR. The extinction terms are deterministic, and no additional measurement-error scatter is injected into the recorded fluxes beyond the per-band stellar-population scatter above. The \simba\ bhALL catalog does not include mass-weighted stellar age or per-galaxy stellar metallicity; the $\tau$-SFH age and Tremonti-relation metallicity are therefore physically motivated proxies, and per-galaxy SPS inputs from the galaxy catalogs are a planned refinement. Broadband colors derive from FSPS spectra evaluated through the LSST bands; the model is an SPS forward model and does not fit a per-galaxy SED.

\subsection{AGN Selection Criteria} \label{sec:agn_selection}

For low-$z$ AGN ($z \lesssim 3$) we impose standard quasar cuts of $g - r < 1.0$, $|r - i| < 2$, and $15 < r < 24$ mag \citep{richards2002}. High-$z$ AGN ($z > 4$) are chosen through Lyman-break criteria \citep{Fan2006,Overzier2006}: $g$-dropouts ($z > 4.0$) require $g - r > 1.2$, $r - i < 0.6$, $i < 25.5$; $r$-dropouts ($z > 5.5$) require $r - i > 1.3$, $i - z < 0.8$, $z < 25.0$.

\subsection{Reference Relation and Regression Target} \label{sec:ml_framework}

We adopt a continuous regression target $\Delta = \log_{10} M_{\rm BH} - \log_{10} M_{\rm BH,med}(M_*, z, {\rm sim})$ in which the reference $\log_{10} M_{\rm BH,med}(M_*, z, {\rm sim})$ is computed separately per simulation as a running median over the same simulation's catalog. The redshift axis is partitioned into bins of width $\Delta z = 0.2$ below $z = 4$, with adaptive widening above $z = 4$ to maintain at least 100 objects per bin so that the high-redshift median is not driven by single-object fluctuations. Within each redshift bin the median $\log_{10} M_{\rm BH}$ is computed in 20 $\log_{10} M_*$ bins between the 1st and 99th percentiles of the stellar mass distribution. The medians are connected with PCHIP monotone shape-preserving interpolation in $\log_{10} M_*$ to suppress spurious oscillation between sparse bins, and linear interpolation is used across redshift. A linear fit to the \simba\ catalog, $\log_{10} M_{\rm BH} = 1.49 \log_{10} M_* - 6.59$, is steeper than both the local relation and the $z = 0$ \simba\ relation $\log_{10} M_{\rm BH} = 1.09 \log_{10} M_* - 3.22$ that \citet{Ma2025} fit to central galaxies with $M_* > 10^{10}~\msun$, motivating a non-parametric per-redshift reference over a fixed global slope; curvature in the underlying scaling and its redshift evolution are absorbed into the reference and kept out of $\Delta$. The running median is computed once from the full catalog, using the same reference within every training fold; at $N \approx 250{,}000$ a single held-out object contributes of order $1\%$ of the objects in its reference bin even in the thinnest high-redshift bins, so the leakage into the reference is small. We impose no class boundary on $\Delta$; the target is the signed offset, so no label-flip filtering is applied.

\subsection{Features, Models, and Cross-Validation} \label{sec:features}

The feature set is photometry-only and contains 27 features: six observed magnitudes in $ugrizy$; six colors ($u-g$, $g-r$, $r-i$, $i-z$, $z-y$, $g-i$); redshift; eight engineered features, namely the SED color curvature $\kappa = (g-r) - 2(r-i) + (i-z)$, the squared color $(g-r)^2$, the spectral range $m_g - m_z$, absolute-magnitude proxies $M_u$, $M_r$, and $M_y$ formed as $m - 5\log_{10}(d_L/10~{\rm pc})$ with $d_L$ the luminosity distance at the catalog redshift and no K-correction, the squared proxy $M_r^2$, and the logarithmic color ratio $\nabla c = \ln[((g-r)+0.5)/((r-i)+0.5)]$; and six cross terms, $(u-g)\,z$, $(g-r)\,z$, $(r-i)\,z$, $(z-y)\,z$, $(g-r)\,m_r$, and $(i-z)\,m_r$. $\lambda_{\rm Edd}$ is not in the feature set because it requires $M_{\rm BH}$.

To attribute the recovery to specific information channels we evaluate a sequence of nested regressors, each adding one information channel to the preceding model: in order of increasing information, the zero-predictor, a regressor given the true $M_*$ and redshift (the true-mass-and-redshift regressor), the host-magnitude proxy, the full-photometry regressor, and its host-only restriction (the host-only regressor); Table~\ref{tab:comparison} tabulates them, and the figure legends use the same descriptive names. The zero-predictor returns the mean, so its scatter is the population scatter $\sigma_{\rm pop}$. Because $\Delta$ is defined as the residual against the per-simulation per-redshift median in $M_*$ (Section~\ref{sec:ml_framework}), a regressor that knows $M_*$ and $z$ exactly should recover nothing beyond the selection-induced skewness signature quantified in Section~\ref{sec:cross_val_prose}, so a near-zero $R^2$ from this regressor is the direct test that the target carries no residual stellar-mass information. The host-magnitude proxy is trained on the host $r$-band magnitude and redshift, the floor that any single broadband flux plus distance can reach. The full-photometry regressor consumes the full 27-feature photometry and is the primary pipeline. The host-only regressor recomputes that feature set from host-only magnitudes with the AGN component zeroed, which with the accretion-rate ablation (Section~\ref{sec:circularity}) separates the host stellar-population channel from the AGN-brightness channel. The first three regressors consume simulation-internal quantities (true stellar mass, or host-only magnitudes) not separately observable for an unresolved AGN-plus-host system, so they are diagnostics; only the full-photometry regressor, built from total magnitudes, corresponds to what a survey measures.

We stack four base regressors, a deep MLP (\code{DeepRegressor}), an FT-Transformer \citep{Gorishniy2021}, a \code{RandomForestRegressor}, and an \code{XGBRegressor}, through a Huber meta-learner with $\delta = 0.5$ (falling back to ordinary least squares if the Huber fit fails to converge). The neural and transformer base models use the same Huber loss to limit the influence of the heavy-tailed $\Delta$ distribution. Five-fold cross-validation is implemented with \code{GroupKFold} keyed on the unique black hole identifier (\code{ID\_bh} in \simba, the SUBFIND ID in \tng\ and \eagle), so that all snapshots of the same black hole fall into the same fold and no near-duplicate temporal copy of a target object can leak from training into evaluation; a runtime assertion verifies the grouping in every fold, and the residual shared-host contamination across folds is below $0.3\%$. Evaluation is done in two modes, ``Magnitude-Limited'' with conservative effective limits ($g, r < 26.0$, $i < 25.5$, $z < 25.0$) chosen to approximate the depths at which extended and blended AGN-plus-host sources remain reliably detected and deblended, and ``Intrinsic'' with no magnitude cuts. These effective limits sit brighter than the nominal LSST ten-year co-added $5\sigma$ point-source depths \citep[$g \approx 27.4$, $r \approx 27.5$, $i \approx 26.8$, $z \approx 26.1$;][]{Ivezic2019}. Cuts in the magnitude-limited mode act on the observed AGN-bearing total fluxes, which preferentially retain over-massive (high-$\Delta$) objects whose AGN light brightens the host, so the magnitude-limited $\sigma_{\rm res}$ is conditional on a $\Delta$-correlated selection function \citep[a selection bias of the kind discussed by][]{Lauer2007}. The primary metric is the residual scatter $\sigma_{\rm res} = {\rm std}(\Delta_{\rm pred} - \Delta_{\rm true})$ over the held-out folds with the mean residual (the bias) removed, reported alongside $R^2$, MAE, and the calibration slope, the slope of an OLS fit of true on predicted $\Delta$, for which unity indicates a unit change in prediction matches a unit change in the true offset. Because $\sigma_{\rm res}$ is mean-subtracted while $R^2$ is computed on the raw residuals, $R^2$ reflects both the scatter and the bias, and the two are not related by $R^2 = 1 - (\sigma_{\rm res}/\sigma_{\rm pop})^2$ when the mean residual is nonzero. The Huber meta-learner pulls predictions toward the conditional median, and since $\Delta$ has median zero but nonzero mean wherever the residual distribution is skewed, a roughly constant mean offset is the expected signature of median-seeking regression on a skewed target. This bias is nearly identical across the nested regressors (Table~\ref{tab:comparison}) and cancels in every increment, so $\sigma_{\rm res}$ is the like-for-like recovery metric for comparisons between models and between modes. We write $\sigma_{\rm pop}$ for the standard deviation of $\Delta$ in the evaluation sample, so that $\sigma_{\rm res} = \sigma_{\rm pop}$ corresponds to no recovery and $\sigma_{\rm res} \to 0$ to full recovery. The intrinsic and magnitude-limited samples are independent seeded $250{,}000$-row draws from different parent pools, with neither nested in the other: the intrinsic sample from the full forward-model catalog, the magnitude-limited sample from the catalog passing the survey-format cuts, which for \simba\ are the low-$z$ AGN selection of Section~\ref{sec:agn_selection} ($z < 4$, $g - r < 1.0$, $|r - i| < 2$, $15 < r < 24$) with an Eddington-ratio floor $\lambda_{\rm Edd} > 0.01$ and the conservative depth limits on top, so the binding cut is $r < 24$; for \tng\ and \eagle\ the depth limits alone. The Eddington-ratio floor selects on $\lambda_{\rm Edd}$, which depends on $M_{\rm BH}$ and hence on $\Delta$, but $\lambda_{\rm Edd}$ is excluded from the feature set because as a feature it would re-encode the target and inflate the apparent recovery, whereas the floor is a selection: it only conditions which objects enter the sample, mirroring that quiescent low-$\lambda_{\rm Edd}$ nuclei are not identified as AGN. The floor is nonetheless a simulation-internal cut, since computing $\lambda_{\rm Edd}$ requires $M_{\rm BH}$; its observable counterpart is a nuclear-luminosity or flux floor, a variant of the selection we do not implement. Recovery within the selected sample is still measured against the true-mass-and-redshift control, which returns $R^2 = -0.009$ in this magnitude-limited sample (Section~\ref{sec:cross_val_prose}). The two differ in redshift reach, population scatter, and unique black hole count (Section~\ref{sec:cross_val_prose}) even though their final row counts nearly coincide; the shortfall below $250{,}000$ is post-draw validity cuts on missing photometry. Increments between the nested regressors (Section~\ref{sec:results}) have bootstrap errors from 500 resamples at a fixed seed. The bootstrap is paired across models: each resample draws the same validation indices for both models of an increment, so the error on the difference reflects the correlation between the two models' residuals. The bootstrap acts on stored out-of-fold residuals, capturing evaluation-set sampling variance but not retraining variance.

\subsection{Cross-Simulation Generalization} \label{sec:cross_sim_methods}

To evaluate whether the learned mapping from photometry to $\Delta$ captures transferable physical content instead of simulation-specific artifacts, we run a transfer experiment in which the regressor is trained on one simulation and evaluated on another. The principal obstacle to transfer is covariate shift: simulations with different feedback and accretion normalizations produce shifted marginal distributions of absolute magnitudes, and tree-ensemble splits learned on the source domain's thresholds therefore land in the wrong place on the target domain even when the color information that carries $\Delta$ is the same. We address this by computing percentile ranks of each photometric feature within adaptive redshift bins on each side, so that the regressor consumes rank features instead of absolute thresholds. This is a domain-adaptation step, and the transfer scatter $\sigma_{\rm res}$ is reported alongside the within-simulation scatter to make the gap explicit. For tests across domains we use the tree-based components (XGBoost, Random Forest) of the stack, since the transformer's attention encodes pairwise feature interactions that tend to be simulation-specific.

\section{Results} \label{sec:results}

\subsection{Simulation Population Properties}

\begin{figure*}[tbp!]
\plotone{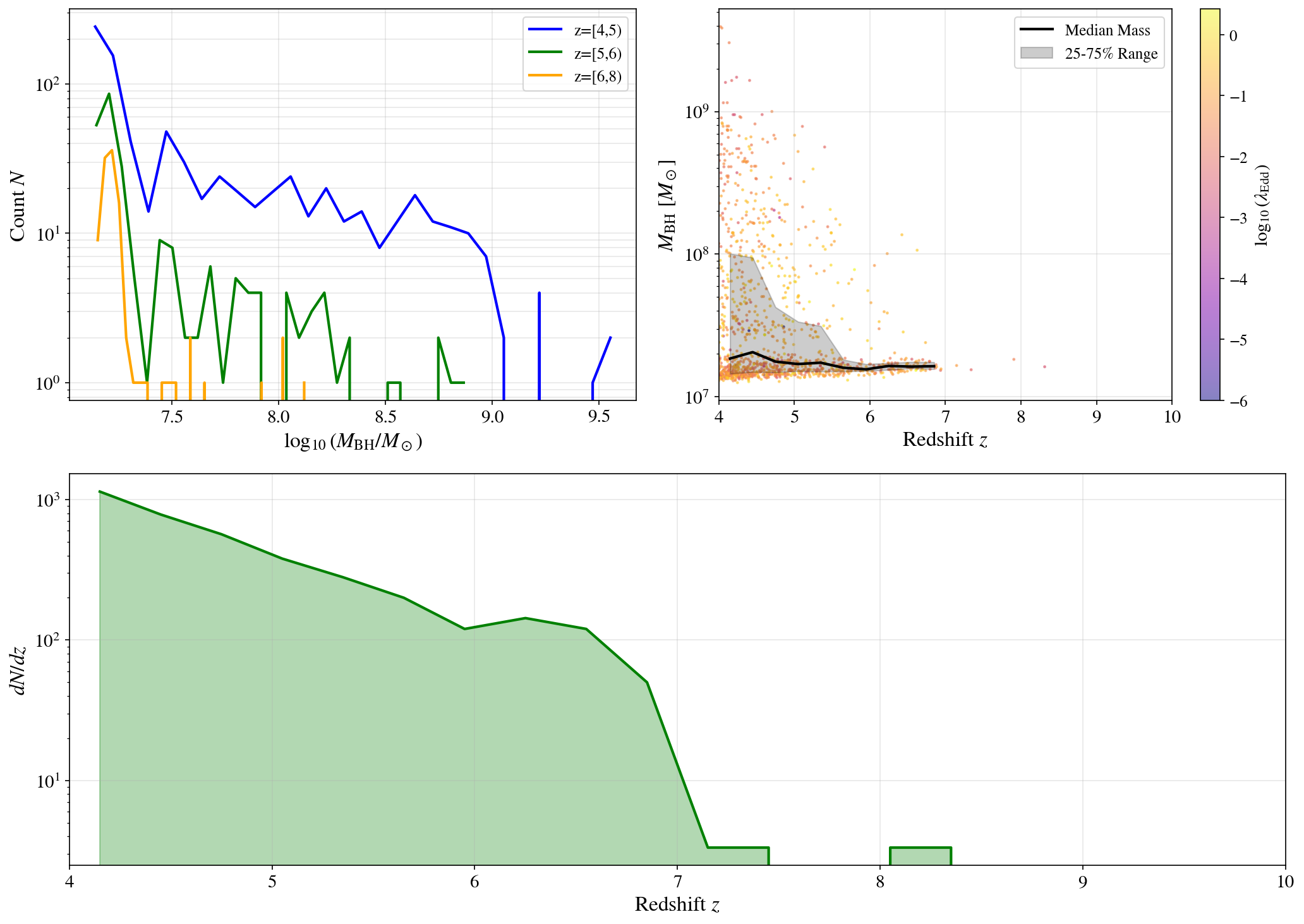}
\caption{Analysis of \simba's black hole population at $z \geq 4$. \textit{Top Left:} Mass function in successive redshift bins, computed from the catalog's black hole particle dynamical masses, whose floor sits at the simulation's mass resolution near $10^{7}~\msun$; the subgrid black hole masses used for the scaling relations and the regression target (Figure~\ref{fig:scaling}) extend down to the seed mass near $10^{4.2}~\msun$. \textit{Top Right:} $M_{\rm BH}$-$z$ relation colored by the Eddington ratio. \textit{Bottom:} Number density evolution.}
\label{fig:simba_analysis}
\end{figure*}

\begin{figure*}[tbp!]
\plotone{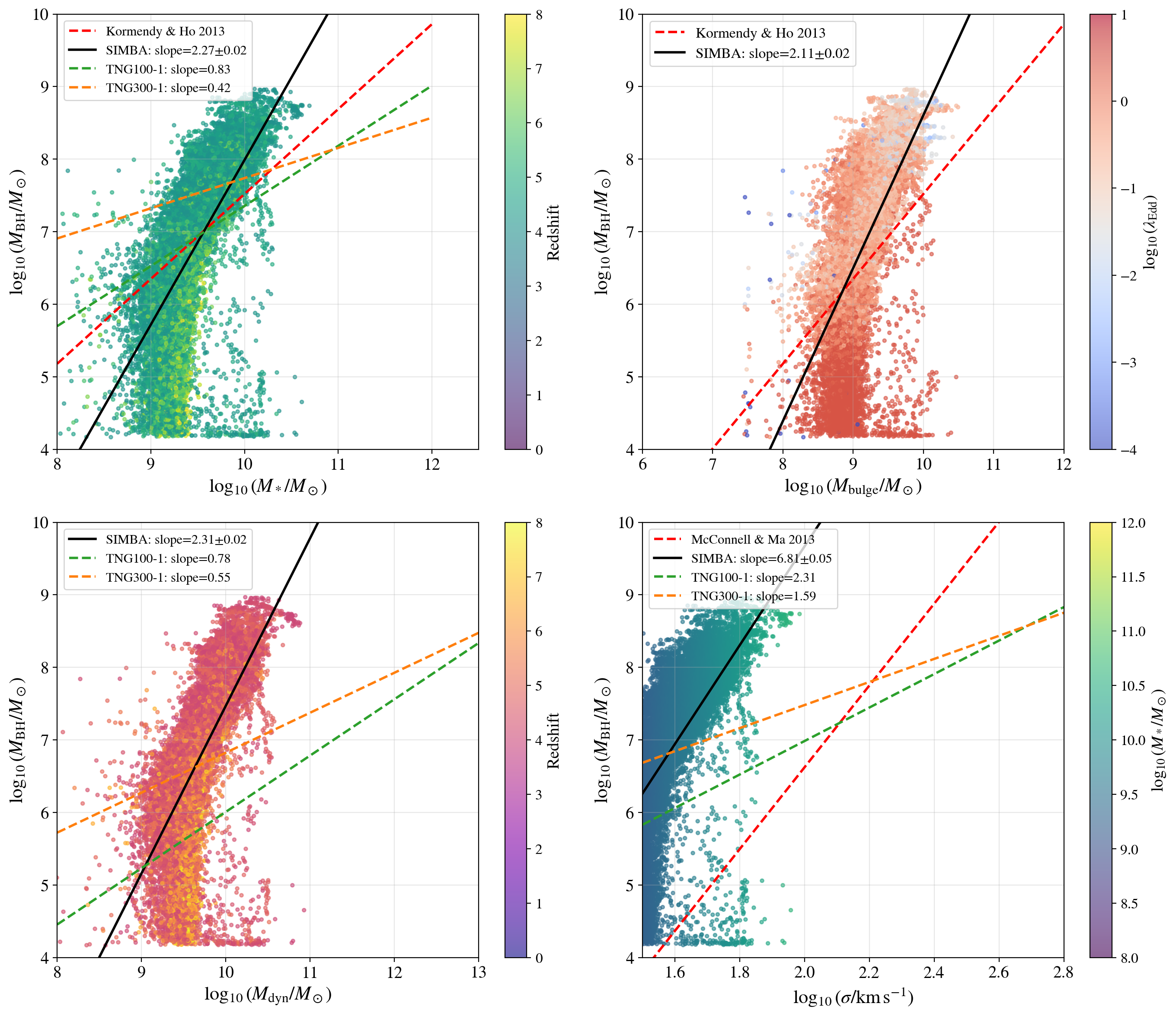}
\caption{\simba\ black hole scaling relations at $z \geq 4$ that follow \citet{Habouzit2021}, with fit slopes of \tng\ overlaid and dashed lines of local scaling relations. The per-redshift running-median reference used for the regression target (Section~\ref{sec:ml_framework}) is fitted independently to each simulation; the linear slopes illustrate the global structure but are not used to define $\Delta$.}
\label{fig:scaling}
\end{figure*}

\begin{figure*}[tbp!]
\plotone{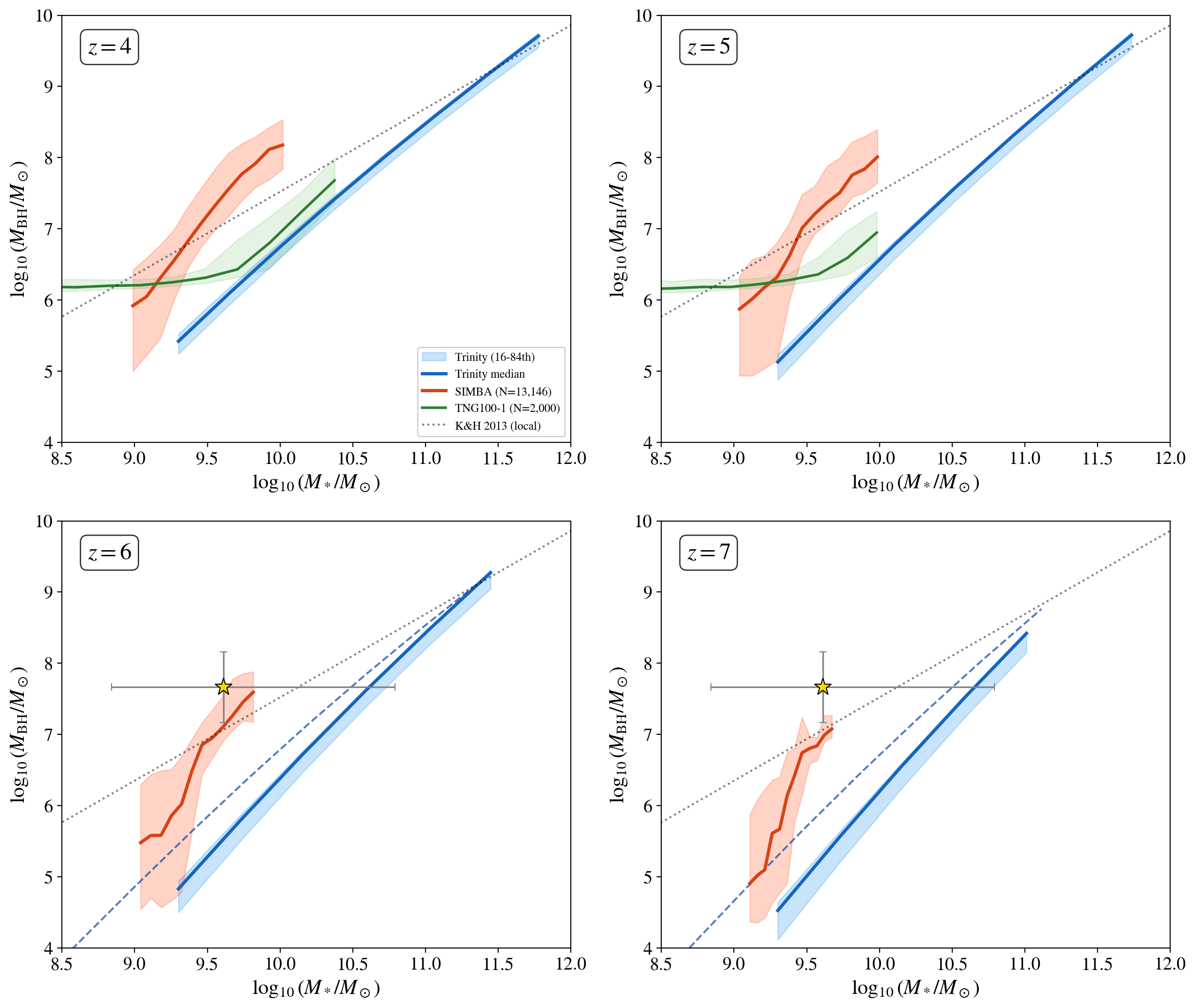}
\caption{Comparison of median $M_{\rm BH}$-$M_*$ relations from \simba\ (red) and \tng\ (green) against the Trinity empirical model (blue; \citealt{Zhang2023Trinity}) at $z = 4$--7. \tng\ is shown at $z = 4$--5, where sufficient black hole growth has occurred for a reliable median relation. The dark blue dashed curve in the $z = 6$ and $z = 7$ panels is Trinity's intrinsic median relation, the underlying relation free of luminosity-selection bias \citep[Fig.~2 of][]{Zhang2024TrinityIV}; the public release tabulates it at $z = 0$ and $z = 6$--10, so it is absent from the $z = 4$ and $z = 5$ panels. Gold stars mark JWST-discovered AGN \citep{Zhang2024TrinityIV}.}
\label{fig:trinity}
\end{figure*}

The trend visible in Figure~\ref{fig:simba_analysis} for \simba's $z \geq 4$ black holes does not clearly support a downsizing interpretation in the redshift range probed. The Eddington ratio median is $\log_{10} \lambda_{\rm Edd} = -1.07$, with a super-Eddington fraction of $1.5\%$ (Figure~\ref{fig:simba_analysis}). The $M_{\rm BH}$--$M_*$ relation slope is $2.27 \pm 0.02$ (Figure~\ref{fig:scaling}), steeper than the \tng\ slopes overlaid in the same figure. This difference is consistent with the distinct accretion prescriptions: \simba's torque-limited model channels cold gas more efficiently onto BHs in gas-rich hosts, producing steeper $M_{\rm BH}$--$M_*$ scaling, while \tng's Bondi-Hoyle accretion with strong kinetic feedback at low Eddington ratios suppresses growth in massive hosts \citep{Habouzit2021,Weinberger2018}. The \simba\ median relation tracks above the Trinity empirical model \citep{Zhang2023Trinity,Zhang2024TrinityIV}, which expects high-redshift, luminosity-selected samples to be biased towards systematically over-massive systems (Figure~\ref{fig:trinity}). \tng\ produces a shallower median $M_{\rm BH}$--$M_*$ relation that sits between \simba\ and the local \citet{KormendyHo2013} scaling at $z = 4$--5, consistent with its more conservative Bondi-limited accretion model.

\subsection{Validation against LSST DP1}

\begin{figure*}[tbp!]
\centering
\includegraphics[width=\textwidth,height=0.92\textheight,keepaspectratio]{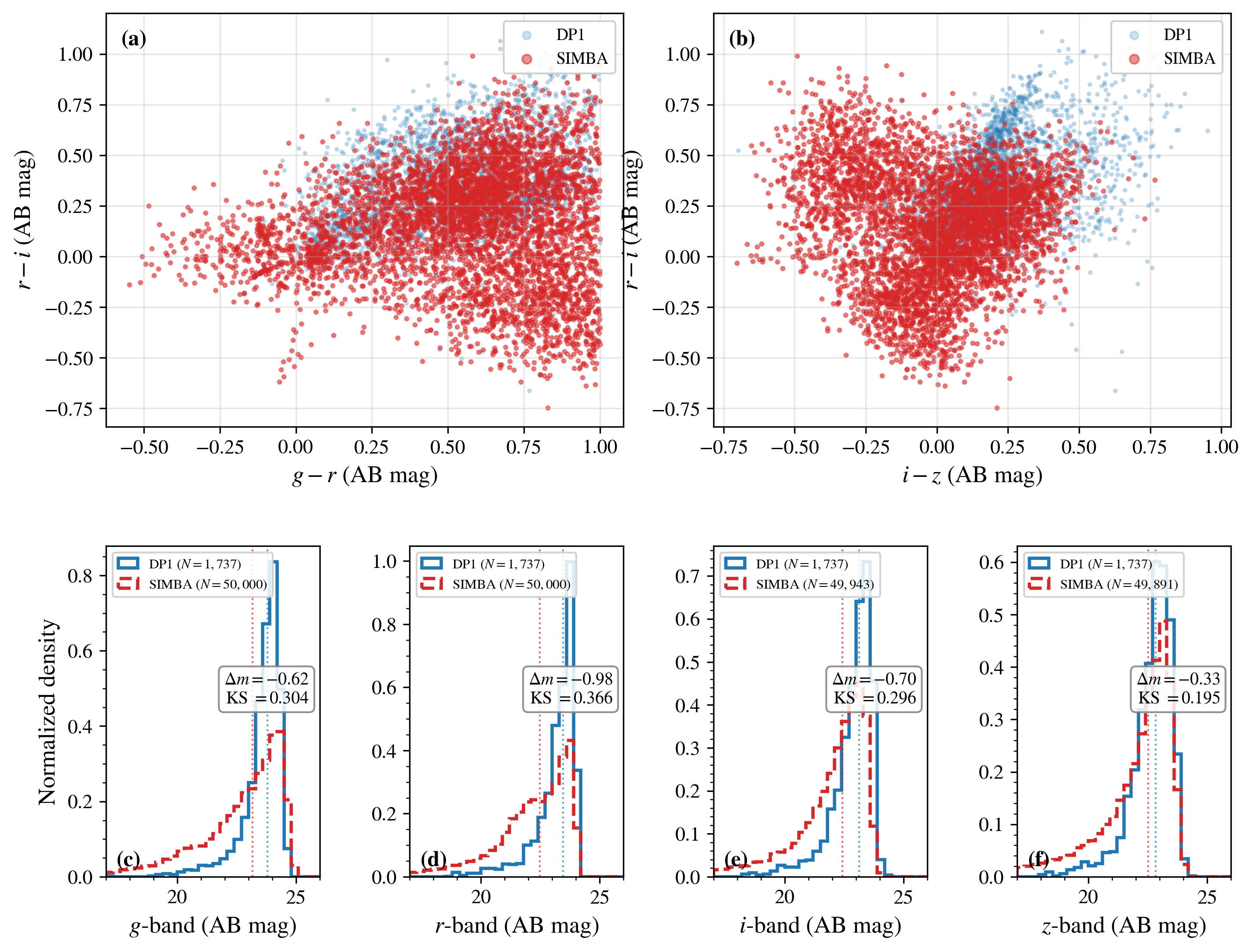}
\caption{Validation of the \simba\ forward model against LSST DP1. Top (a, b): color-color loci ($r-i$ versus $g-r$ and $r-i$ versus $i-z$; \simba\ red, DP1 blue). The narrow tracks in the \simba\ locus are AGN-dominated objects whose unobscured colors are set deterministically by redshift with only a weak luminosity dependence through the ultraviolet slope term (Section~\ref{sec:methods}). At $z < 2$ the median color offsets (\simba\ minus DP1) are $+0.36$ in $g-r$, $-0.28$ in $r-i$, and $-0.37$ in $i-z$, the positive $g-r$ offset indicating the simulated median is redder. Bottom (c--f): band-by-band magnitude distributions at $z < 2$ for $g$, $r$, $i$, $z$, with median offsets $\Delta m = -0.62$, $-0.98$, $-0.70$, $-0.33$ and KS statistics $0.304$, $0.366$, $0.296$, $0.195$. The $z < 2$ DP1 sample is $1{,}737$ objects from the low-$z$ candidate class ($58$ with spectroscopic redshifts, the members of the $439$-object SIMBAD sample of Figure~\ref{fig:dp1_zdist} that fall in this class at $z < 2$; the remaining $1{,}679$ have color-based photometric redshifts from $g-r$ and $r-i$ following the color--redshift method of \citealt{Richards2001photoz}). Over the full low-$z$ candidate class ($N = 2{,}533$; $g - r < 1.0$, no redshift restriction) the cumulative $i$-band medians give an offset of $-0.50$, differing from the $z < 2$ value of $-0.70$ because the samples are defined differently.}
\label{fig:dp1_validation}
\end{figure*}

The forward model reproduces the magnitude ordering of observed DP1 AGN, with the simulated population brighter than DP1 by a few tenths of a magnitude to one magnitude per band and offset in color (Figure~\ref{fig:dp1_validation}). At $z < 2$, the median magnitude offsets (\simba\ median minus DP1 median) are $-0.62$ in $g$, $-0.98$ in $r$, $-0.70$ in $i$, and $-0.33$ in $z$. The corresponding color offsets are $+0.36$ in $g-r$, $-0.28$ in $r-i$, and $-0.37$ in $i-z$, the positive $g-r$ offset indicating that the simulated median is redder than DP1's. The selection cuts and depth limits are applied to the simulated photometry itself, and the recovery measurements compare nested regressors within that same selection, so the model-to-model increments do not depend on the absolute calibration of the forward-modeled photometry against DP1; the DP1 comparison bounds the realism of the forward-modeled catalog and does not enter the recovery measurements. The $r$-band offset shifts the magnitude-limited depth by roughly one magnitude, but the binding $r < 24$ cut sits more than three magnitudes brighter than the LSST ten-year co-added $r \approx 27.5$, so the shifted selection stays within survey reach. Forward-modeled catalogs contain the same mass range observed in high-$z$ quasars \citep{Inayoshi2020}, although the finite $(100~h^{-1}~{\rm cMpc})^3$ simulation volume limits statistics for the rarest objects where $M_{\rm BH} > 10^9~\msun$.

\subsection{Regression Performance} \label{sec:cross_val_prose}

We report recovery on the nested regressors of Section~\ref{sec:features}; the increment between successive regressors, with bootstrap errors, is the quantity of physical interest.

\begin{figure*}[tbp!]
\centering
\includegraphics[width=\textwidth,height=0.92\textheight,keepaspectratio]{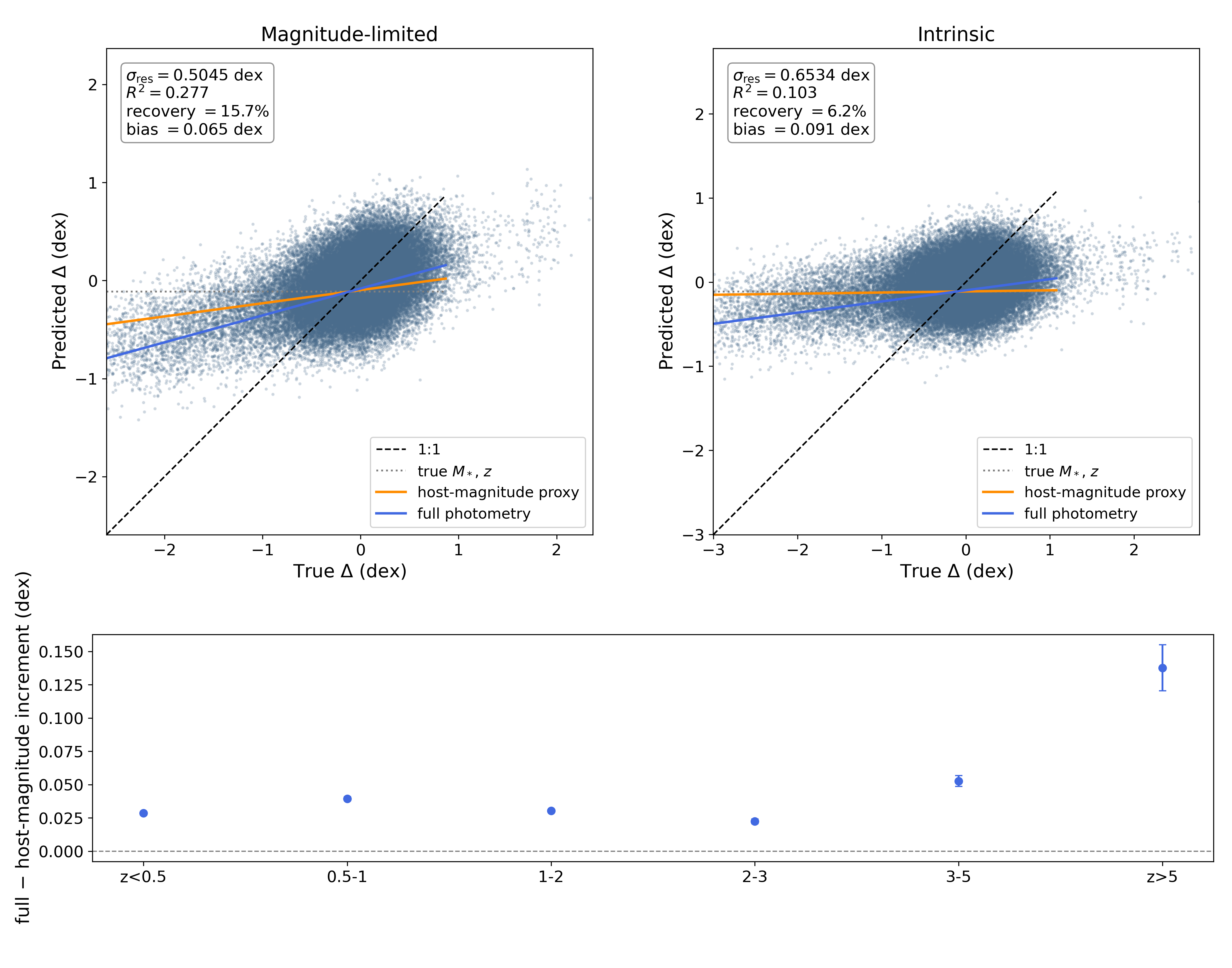}
\caption{\simba\ regression, predicted versus true $\Delta$ over the held-out folds, with the 1:1 line and the ``true $M_*$, $z$'', ``host-magnitude proxy'', and ``full photometry'' trend lines overlaid. Top left (magnitude-limited mode): the full photometry reaches $\sigma_{\rm res} = 0.5045$ dex against a host-magnitude proxy of $0.5588$ dex and a population scatter of $0.5982$ dex on $N = 249{,}439$ samples ($R^2 = 0.277$, recovery $15.7\%$, per-row bias $0.065$ dex). Top right (intrinsic mode): the full photometry reaches $\sigma_{\rm res} = 0.6534$ dex against a host-magnitude proxy of $0.6855$ dex and a population scatter of $0.6966$ dex on $N = 249{,}658$ samples ($R^2 = 0.103$, recovery $6.2\%$, per-row bias $0.091$ dex). Bottom (intrinsic mode): the increment of the full photometry over the host-magnitude proxy in successive redshift bins, whose error bars are independent per-bin paired bootstraps ($500$ resamples of the out-of-fold residuals within each redshift bin). In both scatter panels $R^2$ includes the per-row bias while $\sigma_{\rm res}$ is the bias-removed scatter and the fractional recovery is $1 - \sigma_{\rm res}/\sigma_{\rm pop}$ (Section~\ref{sec:features}).}
\label{fig:ml_regression}
\end{figure*}

For \simba\ at the intrinsic population level ($N = 249{,}658$, $z = 0.00$ to $8.02$, $38{,}280$ unique black holes, five-fold \code{GroupKFold}), the population scatter is $\sigma_{\rm pop} = 0.6966$ dex. The regressor given the true $M_*$ and redshift returns $R^2 = -0.014$; this is the numerical confirmation that the per-simulation per-redshift running-median construction of Section~\ref{sec:ml_framework} orthogonalizes $\Delta$ against stellar mass, leaving $\sigma_{\rm res} \approx \sigma_{\rm pop}$. The orthogonality holds in variance as well as mean: regressing $\Delta^2$, $|\Delta|$, and $\Delta$ on $M_*$ and redshift each returns $R^2 \lesssim 0.02$. The host-magnitude proxy reaches $\sigma_{\rm res} = 0.6855$ dex, and the full photometry reaches $\sigma_{\rm res} = 0.6534$ dex ($R^2 = 0.103$, calibration slope $0.914$; Figure~\ref{fig:ml_regression}), a $6.2\%$ reduction in residual scatter relative to $\sigma_{\rm pop}$. The increment of the full photometry over the host-magnitude proxy is $0.0321 \pm 0.0008$ dex, resolved at $40\sigma$ (statistical); this increment bundles three effects, the broadband colors, the AGN light present in the total magnitudes, and the averaging of per-band intrinsic scatter across six fluxes, so it is not a pure color measurement. The quoted $\sigma$ resolutions reflect the paired-bootstrap statistical error alone, which sits at the percent level; the forward-model systematic from the assumed per-band host-scatter amplitudes shifts the color increments by up to $\sim 15\%$ across a factor-of-four range in scatter amplitude (Section~\ref{sec:limitations}) and dominates; the result rests on the sign and order of magnitude of the recovery. The six raw $ugrizy$ magnitudes plus redshift, with no engineered colors, already reach $\sigma_{\rm res} \approx 0.66$ dex, below the $0.6855$ dex host-magnitude proxy; the engineered colors take the residual the rest of the way to $0.6534$ dex. The stellar-population color increment compares host-only photometry against the host-magnitude proxy, both AGN-free: the host-only regressor ($\sigma_{\rm res} = 0.6612$ dex) over the proxy ($0.6855$ dex) gives $0.6855 - 0.6612 = 0.0243$ dex. The AGN channel is the gain of the full photometry over its host-only restriction, $0.0078 \pm 0.0007$ dex, resolved at $11\sigma$ (statistical). The full-over-single-flux increment varies non-monotonically with redshift, dipping near $z \approx 2$--$3$ before rising above $z = 3$, with per-bin sample sizes from the same evaluation set: $0.0285$ dex at $z < 0.5$ ($N = 87{,}137$), $0.0394$ at $0.5 < z < 1$ ($N = 67{,}181$), $0.0303$ at $1 < z < 2$ ($N = 64{,}911$), $0.0224$ at $2 < z < 3$ ($N = 20{,}755$), $0.0527$ at $3 < z < 5$ ($N = 8{,}731$), and $0.1377$ at $z > 5$ ($N = 943$). Each per-redshift increment bin is resolved above $3\sigma$, including the $z > 5$ bin, which rests on $N = 943$ objects. The color-driven constraint on $\Delta$ therefore strengthens $4.8$-fold from $z < 0.5$ to $z > 5$, toward the epochs where the $M_{\rm BH}$--$M_*$ residual is largest.

In the magnitude-limited mode ($N = 249{,}439$, $z = 0.00$ to $3.73$, $27{,}528$ unique black holes, five-fold \code{GroupKFold}), the population scatter is $\sigma_{\rm pop} = 0.5982$ dex and the regressor given the true $M_*$ and redshift returns $R^2 = -0.009$, again confirming orthogonalization of $\Delta$ against $M_*$. The host-magnitude proxy reaches $\sigma_{\rm res} = 0.5588$ dex and the full photometry reaches $\sigma_{\rm res} = 0.5045$ dex ($R^2 = 0.277$, calibration slope $1.055$; Figure~\ref{fig:ml_regression}), a $15.7\%$ reduction in residual scatter relative to $\sigma_{\rm pop}$ and above the $6.2\%$ recovered at the intrinsic level. The full-over-single-flux increment is $0.0543 \pm 0.0007$ dex, resolved at $78\sigma$ (statistical), the host-only color increment is $0.5588 - 0.5381 = 0.0207$ dex (host-only regressor over the host-magnitude proxy, AGN-free), and the AGN channel is $0.0336 \pm 0.0006$ dex, resolved at $56\sigma$ (statistical), more than four times its intrinsic value of $0.0078$ dex. The survey magnitude cut retains AGN-bright objects whose nuclear light correlates with $M_{\rm BH}$ through accretion, so the AGN channel grows relative to the host-color channel once faint hosts are removed. The full-over-single-flux increment also reverses its redshift trend relative to the intrinsic case: it is largest at $z < 0.5$ ($0.0737$ dex, $N = 127{,}152$) and declines toward higher redshift, $0.0257$ at $0.5 < z < 1$ ($N = 81{,}682$), $0.0094$ at $1 < z < 2$ ($N = 36{,}946$), and $0.0122$ at $2 < z < 3$ ($N = 3{,}569$); these bins sum to $249{,}349$ of the $249{,}439$ samples, with the remaining 90 objects at $z > 3$ falling below the 200-object threshold for a tabulated bin. The magnitude cut removes faint hosts at higher redshift and leaves AGN-dominated systems whose colors encode less stellar-population information, the opposite of the intrinsic sample where the color increment strengthens with redshift.

The \eagle\ volumes follow the same nested comparison. RefL0025 recovers a $4.8\%$ reduction in residual scatter at the intrinsic level ($N = 25{,}135$, $\sigma_{\rm pop} = 0.1805$ to $\sigma_{\rm res} = 0.1719$ dex) with an intrinsic AGN channel consistent with zero, its host stellar populations carrying nearly the whole recoverable signal; the flagship RefL0100 volume, at roughly ten times the sample ($N = 254{,}199$), recovers $11.3\%$ ($0.2010$ to $0.1783$ dex) and turns that intrinsic AGN channel into a small but positive one while keeping the host-leaning ordering. Both volumes run in both modes (Table~\ref{tab:comparison}); the heavy \eagle\ seed mass of $1.475 \times 10^5~\msun$ and thermal-feedback prescription are consistent with the weak coupling between AGN luminosity and $\Delta$, though neither is isolated as its cause (Section~\ref{sec:circularity}).

The \tng\ volumes likewise follow the same nested regressors (Table~\ref{tab:comparison}). TNG100-1 recovers the largest fractional reduction of any simulation in the primary analysis, $17.4\%$ at the intrinsic level ($N = 15{,}242$, $\sigma_{\rm pop} = 0.1942$ to $\sigma_{\rm res} = 0.1604$ dex), and TNG300-1 recovers $9.7\%$ intrinsic ($N = 28{,}074$, $0.1900$ to $0.1716$ dex); both run in both modes. Across these \tng\ and \eagle\ volumes the regressor given the true $M_*$ and redshift returns $R^2 \in [-0.012, +0.031]$. Per-volume channel splits are dissected in Section~\ref{sec:circularity}.

The small positive values trace to the interaction of the reference construction with each sample's residual distribution: the running median zeroes the conditional median of $\Delta$ at fixed $(M_*, z)$ within each evaluation sample, while a squared-error regressor recovers the conditional mean, which is nonzero wherever the residual distribution is skewed (Section~\ref{sec:features}). In the \eagle\ RefL0100 magnitude-limited sample, where the value is largest, two contributions produce this skew: the heavy seed piles low-mass objects at the low-$\Delta$ boundary, a pileup the $\Delta$-correlated survey retention reshapes across the mass range, and the reference extrapolates flat beyond its 99th-percentile stellar-mass support. Re-centering $\Delta$ on the conditional mean and excluding $\log_{10} M_* > 11$ together collapse the value from $+0.031$ to $+0.003$, so the median orthogonalization holds as constructed and this residual $R^2$ measures distribution shape and contains no recoverable mass information. The pattern across simulations follows the seed prescriptions: \simba's light seed populates a nearly uniform under-massive tail whose constant conditional mean yields no recovery ($R^2 = -0.014$ and $-0.009$) despite the largest global skew, the well-resolved \tng\ volumes return $+0.002$ to $+0.007$, and the heavy-seeded \eagle\ volumes under the survey selection, together with the low-resolution \tng\ variants (Appendix~\ref{sec:appendix_resolution}), show the selection-shaped, $M_*$-graded skew that a mean-recovering regressor reads as a small positive $R^2$.

\begin{figure*}[tbp!]
\centering
\includegraphics[width=\textwidth]{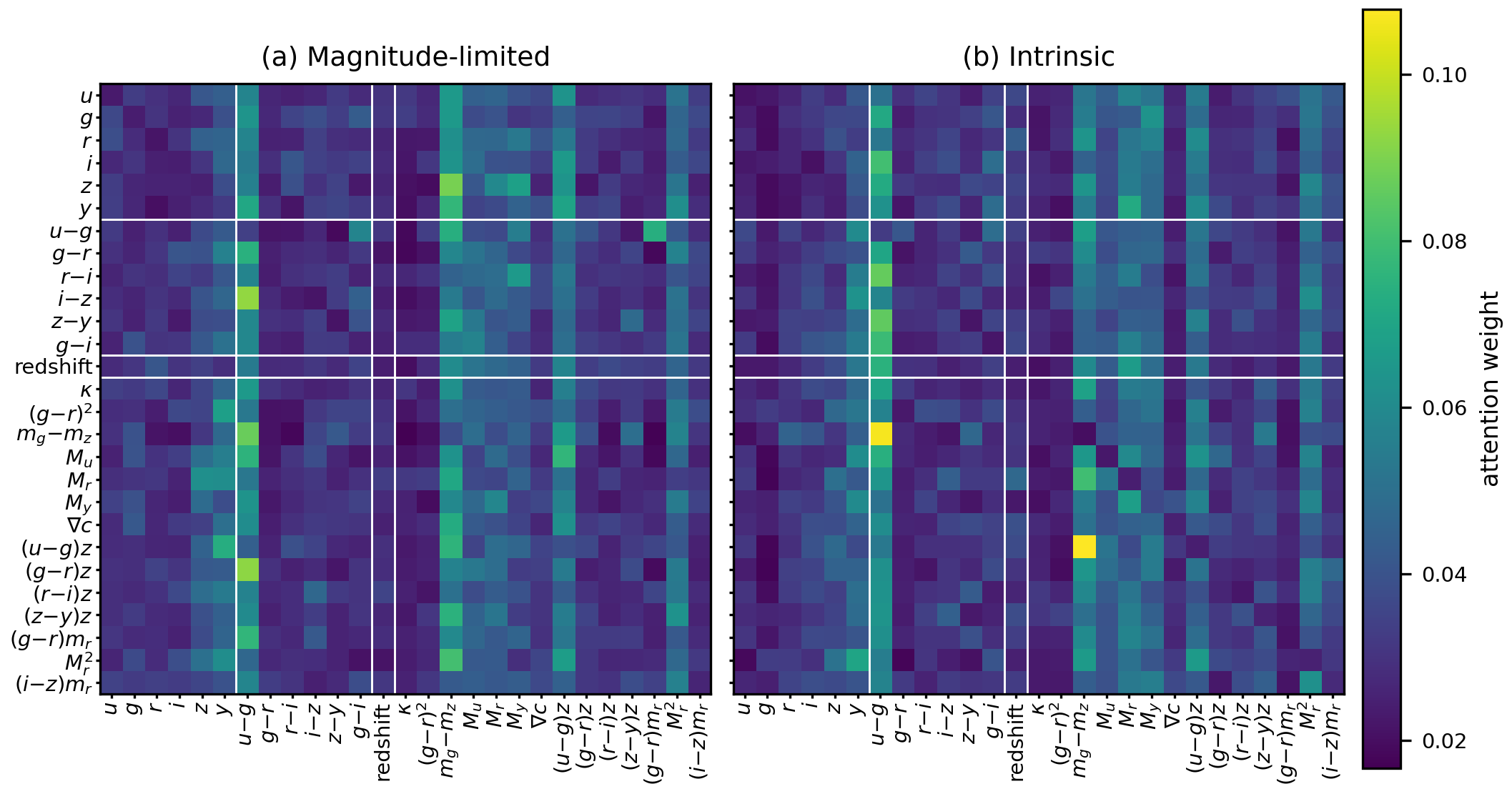}
\caption{FT-Transformer attention heatmaps for \simba\ in the magnitude-limited mode (panel a) and the intrinsic mode (panel b), computed over the 27-feature photometry set used in the regressor and shown on a shared color scale. Rows and columns give the feature names, with thin white rules separating the apparent magnitudes, the colors, redshift, and the engineered features and cross terms; $\lambda_{\rm Edd}$ is not in the feature set and does not appear in these panels.}
\label{fig:attention_heatmaps}
\end{figure*}

The FT-Transformer attention heatmaps (Figure~\ref{fig:attention_heatmaps}) concentrate on the absolute-magnitude proxy, the spectral range $m_g - m_z$, and the $u-g$ color, consistent with reliance on broad SED shape and the host stellar-population channel.

\subsection{Signal Decomposition} \label{sec:circularity}

\begin{figure}[tbp!]
\centering
\includegraphics[width=\columnwidth,height=0.42\textheight,keepaspectratio]{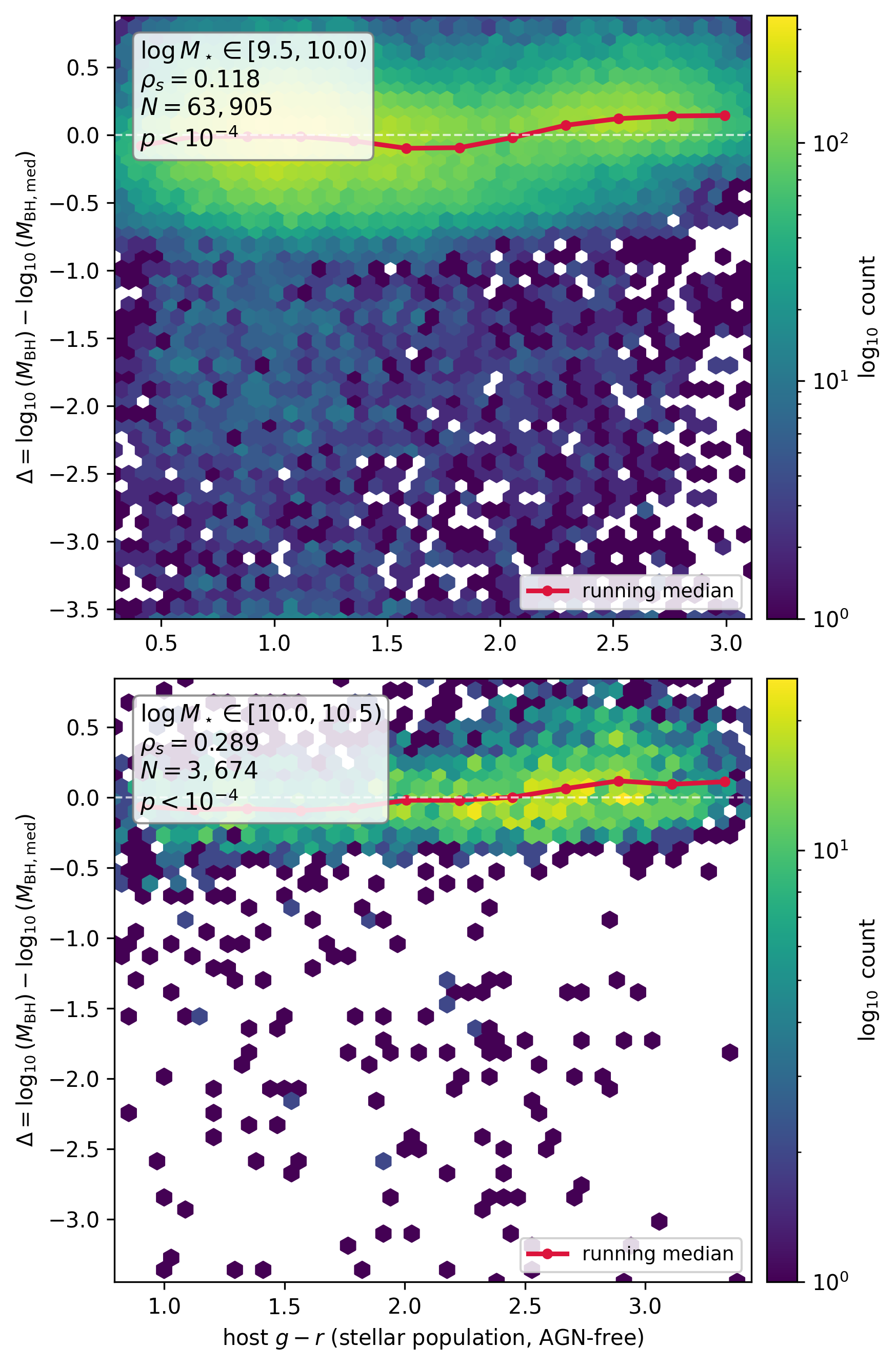}
\caption{Residual $\Delta$ against host-only $g-r$ color in two fixed stellar-mass bins ($\log_{10} M_* = 9.5$--$10$, $10$--$10.5$) for \simba\ at $z < 1.5$, restricted to objects within the reference's per-redshift PCHIP stellar-mass support. In both bins redder hosts at fixed $M_*$ are more over-massive, with Spearman $\rho = +0.118$ ($N = 63{,}905$) and $+0.289$ ($N = 3{,}674$), both at $p < 10^{-4}$. Resolving each bin by color locates the coupling in the heavily attenuated $g-r \geq 2$ tail (Section~\ref{sec:circularity}); it routes through the dust column.}
\label{fig:drivers}
\end{figure}

Because each simulation's accretion prescription couples $\dot{M}$ to $M_{\rm BH}$ and to the nuclear gas conditions, and the forward model then links $\dot{M} \to L_{\rm bol} \to$ magnitudes, the photometric information about $\Delta$ can flow through two distinct channels: the host stellar populations, whose colors track specific star formation rate and stellar age, and the AGN brightness, set by the accretion rate. The nested regressors of Section~\ref{sec:features} separate these. The stellar-population channel is the gain of the host-only regressor over the no-recovery floor, the $\sigma_{\rm pop}$ scatter reached by a regressor given the true $M_*$ and redshift: for \simba\ intrinsic samples this is $0.6966 \to 0.6612$ dex, about $0.035$ dex; host colors carry most of the photometric information about $\Delta$. Because the host colors are generated from $(M_*, {\rm SFR}, z)$ through the FSPS recipe, this channel is bounded by the information those catalog quantities re-encode into the SED, so it is an upper bound on what color adds beyond stellar mass and star formation rate and does not measure independent information in observed galaxy colors. The AGN-brightness channel is measured two ways: as the gain of the full photometry over its host-only restriction, $0.0078 \pm 0.0007$ dex, and by the accretion-rate ablation, which randomly permutes $\dot{M}$ at fixed redshift to break the $M_{\rm BH} \to L_{\rm bol}$ correlation while leaving the host SED intact; its difference from the unshuffled full photometry isolates the same AGN channel through the forward model directly. Shuffling $\dot{M}$ raises the full-photometry scatter from $0.6534$ to $0.6710$ dex ($N = 249{,}477$). That rise decomposes into two terms: the AGN information channel, equal to the feature-restriction measure ($+0.0078$ dex), and a contamination cost of $+0.0098$ dex from unpredictable nuclear flux corrupting the host colors in the shuffled catalog. The AGN information channel is therefore the $+0.0078$ dex feature-restriction value, and the larger $+0.0176$ dex shuffle drop adds a contamination artifact from the shuffled catalog and overstates it. Two consistency checks hold: the shuffled-catalog regressor given the true $M_*$ and redshift lands at $0.6951$ dex against the main floor $0.6966$ dex, and the shuffled host-only regressor ($0.6623$ dex) reproduces the main host-only value ($0.6612$ dex) to within $0.0011$ dex, as the host features are untouched. The stellar-population channel therefore dominates the \simba\ intrinsic signal on three independent measurements: feature restriction, causal ablation, and the host color-$\Delta$ coupling of Figure~\ref{fig:drivers}.
The two channels exchange weight under the survey selection. In the magnitude-limited sample the stellar-population channel, the host-only regressor over the no-recovery floor, is $0.5982 \to 0.5381$ dex, about $0.060$ dex, and the AGN channel grows to $0.0336 \pm 0.0006$ dex, more than four times its intrinsic value, while the host-only color increment over the host-magnitude proxy falls to $0.0207$ dex.

The ordering of the AGN channel across simulations is consistent with the nominal black hole mass dependence of their accretion prescriptions, but the prescriptions do not fix it. In every simulation the tabulated $\dot{M}$ depends not only on $M_{\rm BH}$ but on the density and sound speed of the gas around the black hole, on its angular momentum in \eagle\ \citep{RosasGuevara2015}, and on the disk fraction and enclosed mass within the black hole kernel in \simba\ \citep{AnglesAlcazar2017}; none of these is in the catalogs, and all differ substantially from galaxy to galaxy and between simulations whose feedback regulates the nuclear gas differently. The formal mass scalings only bracket the range: \simba's cold-gas torque-limited mode scales as $\dot{M} \propto M_{\rm BH}^{1/6}$ \citep{Hopkins2011}, nearly mass-independent, with its hot-gas Bondi mode contributing where the ambient gas is hot, whereas the Bondi-Hoyle rates of \tng\ and \eagle\ scale as $M_{\rm BH}^2$ at fixed gas conditions \citep{Weinberger2017,RosasGuevara2015}, so a steeper coupling of nuclear luminosity to black hole mass, and with it a larger AGN channel, is plausible in the Bondi-based simulations. The measured channels follow this expectation only in part. Three of the four \tng\ datapoints are AGN-leaning, the AGN channel exceeding the host-only color increment in both TNG300-1 modes and most strongly in TNG100-1 intrinsic, the reverse of the \simba\ intrinsic ordering; in the TNG100-1 magnitude-limited mode the two channels are comparable and the host-color channel is marginally larger, so the AGN-leaning signature is present in the intrinsic samples but mixed under the survey cut at this sample size ($N = 2{,}232$). \eagle, whose prescription shares \tng's nominal $M_{\rm BH}^2$ scaling, nevertheless sits at the host-leaning end, with an intrinsic AGN channel three times smaller than TNG100-1's ($0.0074$ against $0.0230$ dex in the flagship volume) and consistent with zero in RefL0025, so the nominal mass exponent does not by itself set the channel's amplitude; the gas conditions and feedback histories that also enter $\dot{M}$ evidently do. The three simulations thus span a range of channel weightings, \simba\ and \eagle\ host-leaning at the intrinsic level, \tng\ AGN-leaning, and \simba\ shifting toward AGN-leaning under the survey cut, which we read as a property of each simulation's accretion and feedback implementation as a whole rather than of its accretion formula alone. Because \simba's magnitude-limited selection is the low-$z$ AGN cut (Section~\ref{sec:agn_selection}) with an Eddington-ratio floor while the \tng\ and \eagle\ magnitude-limited samples apply depth limits alone, cross-simulation comparisons of magnitude-limited channel weights are between non-parallel selections. The ordering across simulations rests on the intrinsic modes, which are selected identically.

The flagship statistics refine the small-volume \eagle\ picture, host-leaning in both volumes. In RefL0025 the host-only photometry reproduces the full photometry to within $0.0005 \pm 0.0005$ dex in the intrinsic mode and $0.0023 \pm 0.0007$ dex in the magnitude-limited mode, so the AGN channel is consistent with zero in the intrinsic population and only marginally nonzero under the survey cut. The flagship RefL0100 volume, with roughly ten times the sample in each mode, resolves a small but positive AGN channel, $0.0074 \pm 0.0006$ dex in the intrinsic mode and $0.0072 \pm 0.0016$ dex in the magnitude-limited mode, while the host-color channel remains the larger of the two ($0.0138$ and $0.0120$ dex respectively). \eagle's thermal-feedback prescription and heavy seed mass are plausible contributors to the weak coupling between AGN luminosity and $\Delta$; the nuclear gas conditions that also set $\dot{M}$ are not tabulated, so we do not attribute it to either ingredient alone.

The physical origin of the stellar-population channel is the catalog-level correlation between the $M_{\rm BH}$--$M_*$ residual and host star formation history at fixed stellar mass. Figure~\ref{fig:drivers} shows $\Delta$ against host-only $g-r$ color in two fixed $\log_{10} M_*$ bins at $z < 1.5$, restricted to objects within the reference's per-redshift PCHIP stellar-mass support so the flat extrapolation beyond that support does not enter (Section~\ref{sec:cross_val_prose}): in each bin a redder host at fixed $M_*$ corresponds to a more over-massive black hole, with Spearman $\rho = +0.118$ ($N = 63{,}905$) and $+0.289$ ($N = 3{,}674$), both at $p < 10^{-4}$. The upper bin rises from the $+0.130$ measured over its full extent because $51.2\%$ of that bin lay beyond the stellar-mass support, where the flat extrapolation inflates $\Delta$ and dilutes the coupling (Section~\ref{sec:cross_val_prose}); the former $\log_{10} M_* = 10.5$--$11$ bin lay entirely beyond support and is dropped. The sign matches the catalog-level $\Delta$--sSFR correlation, in which more over-massive systems formed their stars earlier and are now more quiescent at fixed $M_*$ \citep{thomas19,Ma2025,martin18}, consistent with quiescence tracking black hole mass at fixed stellar mass \citep{Terrazas2016,Terrazas2020,Davies2020}. Resolving each bin by color locates the coupling: blueward of $g-r = 2$ the bulk shows no consistent positive signal ($\rho = -0.049$ in the lower bin, $+0.093$ in the upper), while the heavily attenuated tail at $g-r \geq 2$ carries it ($\rho = +0.186$ and $+0.194$, holding $26.5\%$ and $66.7\%$ of each bin). The SFR- and gas-fraction-dependent \citet{Calzetti2000} attenuation drives the trend: the positive coupling is confined to the reddened tail where the attenuation term sets the color, and is absent from the blueward bulk where color tracks the stellar-age sequence directly, so the star-formation-history information the $\Delta$--sSFR correlation encodes reaches broadband color through the host dust column. Rest-frame band mixing across $0 < z < 1.5$, where the observed $g-r$ samples different rest wavelengths at different redshifts, is a limitation of the fixed-band analysis. The colors in this figure are generated from $(M_*, {\rm SFR}, z)$ by construction, so the figure does not independently establish the catalog-level correlation between $\Delta$ and specific star formation rate, whose existence rests on the simulation itself \citep{thomas19,Ma2025}; it tests whether that coupling survives transmission through the SED, dust, intrinsic scatter, and band-mixing chain into broadband colors, and the measured amplitude is conditional on the adopted star formation history proxy and scatter values. This is the relationship the host-only regressor exploits to improve on the host-magnitude proxy.

\subsection{Regression Results Across Simulations} \label{sec:cross_val}

The nested regressors are applied identically across simulations and resolutions, and Table~\ref{tab:comparison} collects every regressor, mode, and increment with bootstrap errors, interpreted in Sections~\ref{sec:cross_val_prose} and~\ref{sec:circularity}. The \eagle\ simulation suite is detailed in Appendix~\ref{sec:appendix_eagle}; the low-resolution TNG100-2 and TNG300-2 runs are excluded from primary results as resolution-limited and reported in Appendix~\ref{sec:appendix_resolution} (Table~\ref{tab:resolution_artifact}).

\begin{table*}[tp!]
\centering
\caption{Photometric Recovery of the Mass Offset Across Simulations \label{tab:comparison}}
\footnotesize
\setlength{\tabcolsep}{3pt}
\begin{tabular}{llrcccccc}
\toprule
\textbf{Simulation} & \textbf{Mode} & \textbf{$N$} & \textbf{$\sigma_{\rm pop}$} & \textbf{$R^2$} & \textbf{$\sigma_{\rm res}$} & \textbf{$\sigma_{\rm res}$} & \textbf{$\sigma_{\rm res}$} & \textbf{$R^2$} \\
 & & & \textbf{(dex)} & \textbf{(true $M_*$, $z$)} & \textbf{host mag (dex)} & \textbf{full phot. (dex)} & \textbf{host-only (dex)} & \textbf{full phot.} \\
\hline
\simba & Intrinsic & 249,658 & $0.6966$ & $-0.014$ & $0.6855$ & $0.6534$ & $0.6612$ & $0.103$ \\
\simba & Magnitude-Limited & 249,439 & $0.5982$ & $-0.009$ & $0.5588$ & $0.5045$ & $0.5381$ & $0.277$ \\
TNG100-1 & Intrinsic & 15,242 & $0.1942$ & $+0.004$ & $0.1911$ & $0.1604$ & $0.1833$ & $0.318$ \\
TNG100-1 & Magnitude-Limited & 2,232 & $0.2193$ & $+0.007$ & $0.2161$ & $0.2036$ & $0.2085$ & $0.137$ \\
TNG300-1 & Intrinsic & 28,074 & $0.1900$ & $+0.006$ & $0.1889$ & $0.1716$ & $0.1835$ & $0.182$ \\
TNG300-1 & Magnitude-Limited & 3,709 & $0.240$ & $+0.002$ & $0.239$ & $0.224$ & $0.236$ & $0.119$ \\
\eaglesm & Intrinsic & 25,135 & $0.1805$ & $-0.012$ & $0.1804$ & $0.1719$ & $0.1723$ & $0.054$ \\
\eaglesm & Magnitude-Limited & 2,649 & $0.2693$ & $+0.017$ & $0.2687$ & $0.2662$ & $0.2685$ & $0.022$ \\
\eagle\ RefL0100 & Intrinsic & 254,199 & $0.2010$ & $+0.012$ & $0.1995$ & $0.1783$ & $0.1857$ & $0.186$ \\
\eagle\ RefL0100 & Magnitude-Limited & 25,256 & $0.3043$ & $+0.031$ & $0.2987$ & $0.2795$ & $0.2867$ & $0.154$ \\
\hline
\bottomrule
\end{tabular}
\tablecomments{Columns: $\sigma_{\rm pop}$ is the population scatter of $\Delta$, the scatter of the zero-predictor; $R^2$ (true $M_*$, $z$) is the coefficient of determination of the regressor given the true stellar mass and redshift, expected near zero by construction (Section~\ref{sec:cross_val_prose}); the host mag, full phot., and host-only columns list $\sigma_{\rm res}$, the mean-subtracted residual standard deviation, for the host-magnitude proxy, the full 27-feature photometry, and the host-only regressor; the final column lists the full photometry's $R^2$. Increments (dex; bootstrap errors from 500 resamples) are listed per simulation and mode as full photometry vs host-magnitude proxy (colors $+$ AGN $+$ scatter averaging); host-only vs host-magnitude proxy (host colors, AGN-free); full photometry vs host-only (AGN). \simba\ intrinsic: $0.0321 \pm 0.0008$; $0.0243$; $0.0078 \pm 0.0007$. \simba\ magnitude-limited: $0.0543 \pm 0.0007$; $0.0207$; $0.0336 \pm 0.0006$. TNG300-1 intrinsic: $0.0174 \pm 0.0006$; $0.0054$; $0.0119 \pm 0.0006$. TNG300-1 magnitude-limited: $0.014$; $0.003$; $0.012$. The TNG300-1 magnitude-limited entries are quoted at three decimals and without bootstrap errors because that row's per-object out-of-fold residuals were not persisted, so the paired bootstrap could not be run and the values are reported at the precision recorded in the production log. TNG100-1 intrinsic: $0.0308 \pm 0.0021$; $0.0078$; $0.0230 \pm 0.0019$. TNG100-1 magnitude-limited: $0.0125 \pm 0.0017$; $0.0076$; $0.0049 \pm 0.0015$. \eaglesm\ intrinsic: $0.0085 \pm 0.0006$; $0.0081$; $0.0005 \pm 0.0005$. \eaglesm\ magnitude-limited: $0.0025 \pm 0.0008$; $0.0002$; $0.0023 \pm 0.0007$. \eagle\ RefL0100 intrinsic: $0.0212 \pm 0.0006$; $0.0138$; $0.0074 \pm 0.0006$. \eagle\ RefL0100 magnitude-limited: $0.0192 \pm 0.0024$; $0.0120$; $0.0072 \pm 0.0016$. The reported $R^2$ is computed on raw residuals and so includes a per-row bias that differs from model to model by only $\sim 0.01$ dex; it cancels in every increment, so $\sigma_{\rm res}$ is the like-for-like recovery metric (Section~\ref{sec:features}). \simba's magnitude-limited row applies the low-$z$ AGN color selection with the $\lambda_{\rm Edd} > 0.01$ floor, whereas the \tng\ and \eagle\ magnitude-limited rows apply depth limits alone; the magnitude-limited rows are therefore not a parallel cross-simulation selection, whereas the intrinsic rows are.}
\end{table*}

\subsection{Cross-Simulation Transfer} \label{sec:cross_sim_results}

Cross-simulation transfer experiments are presented in Table~\ref{tab:cross_sim} for the intrinsic mode, with six directions spanning the three simulation pairs (\simba\ $\leftrightarrow$ TNG100-1, \simba\ $\leftrightarrow$ \eaglesm, TNG100-1 $\leftrightarrow$ \eaglesm). Each direction trains the tree-based components of the stack on the source simulation and evaluates the residual scatter $\sigma_{\rm res}$ on the target, reported alongside the within-simulation scatter of the same XGBoost-plus-Random-Forest ensemble and the target's population scatter $\sigma_{\rm pop}$. Transfer degradation is read against the like-for-like XGBoost-plus-Random-Forest baseline in Table~\ref{tab:cross_sim}, since the transfer runs use only the tree components; full-ensemble within-simulation values are in Table~\ref{tab:comparison}.

The cross-simulation transfer experiments run on a separate matched subsample of each catalog; the \simba\ population scatter for this subsample ($0.692$ dex) differs slightly from the $0.6966$ dex of the full-sample primary analysis. Because $\sigma_{\rm pop}$ differs by roughly a factor of three between \simba\ and the narrower TNG100-1 and \eaglesm\ targets, degradation should be read relative to each target's $\sigma_{\rm pop}$. Transfer into \simba\ degrades only slightly: a model trained on TNG100-1 or \eaglesm\ and evaluated on \simba\ recovers $\sigma_{\rm res} = 0.690$ to $0.695$ dex against a within-simulation baseline of $0.668$ dex, a degradation of $0.022$ to $0.027$ dex, about $3$ to $4$ percent of $\sigma_{\rm pop}$. Transfer out of \simba\ into the narrower targets is large: $\sigma_{\rm res} = 0.393$ to $0.521$ dex against within-simulation baselines of $0.163$ to $0.171$ dex, a degradation of $0.230$ to $0.350$ dex, which exceeds the entire population scatter of the target. The transfer between the two narrow targets is intermediate and direction-dependent, $0.017$ dex for TNG100-1 $\to$ \eaglesm\ but $0.132$ to $0.133$ dex for \eaglesm\ $\to$ TNG100-1. A model trained on \simba's broad $\Delta$ distribution transfers its predictions into the narrower-distribution targets poorly because the residual structure it learns is in part specific to the \simba\ accretion and feedback prescription; the wide $\sigma_{\rm pop}$ of \simba\ as a target, conversely, leaves room for a foreign model to land close without resolving the fine structure. This out-of-\simba\ scatter of $0.39$ to $0.52$ dex exceeds the target populations' total scatter of $0.18$ to $0.19$ dex, a negative $R^2$: a \simba-trained model deployed on a TNG-like or \eagle-like universe is anti-informative relative to predicting the population mean.

Raw and rank-normalized configurations perform nearly identically under the per-simulation reference and the continuous regression target: across all six intrinsic directions the raw-versus-rank difference in $\sigma_{\rm res}$ is at most $0.014$ dex and $0.003$ dex or less in every other direction. Under a single global reference, rank normalization compensates for the absolute-magnitude offset between simulations; the per-simulation reference removes that offset, so raw and rank features perform identically.

The failure of transfer decomposes into three layers that separate a scale artifact from an information limit. First, the raw correlation between a transferred prediction and the target $\Delta$ is weak in every direction: the maximum cross-simulation Pearson $r$ is $0.155$ (\simba\ $\to$ TNG100-1, rank), and all others fall below it. Second, the residual after a blind moment-matched calibration, which rescales the transferred prediction to the target population mean and variance without using any labeled target points, is $\sigma_{\rm cal} = \sigma_{\rm pop}\sqrt{2(1-r)}$, so calibration beats the population baseline only when $r > 0.5$; at the measured $r \leq 0.155$ the calibrated scatter exceeds $\sigma_{\rm pop}$ in every direction. Third, a leave-one-simulation-out test that trains on the pooled other two simulations and evaluates on the held-out one reaches Pearson $r \leq 0.165$ and likewise calibrates above $\sigma_{\rm pop}$ in all entries, so pooling simulations does not recover transferable structure. The one direction with a small raw degradation, TNG100-1 $\to$ \eaglesm\ at $0.017$ dex, is an artifact: its prediction is near-constant on the target (Spearman $\rho \approx -0.004$), which yields a small scatter against a narrow target distribution.

Conditional residualization isolates the component of the color-$\Delta$ coupling that is invariant to the marginal distribution shifts between simulations. For each dataset we form conditional color offsets $\delta_c = c - {\rm median}(c \mid {\rm abs\text{-}mag\ decile} \times z\ {\rm bin})$, computed in cells of at least $100$ objects on the full intrinsic frames, and measure the rank correlation of $\delta_c$ with $\Delta$. The conditional couplings are listed in Table~\ref{tab:couplings}. Their sign is universal across the three simulations: at fixed magnitude and redshift, over-massive black holes sit in bluer hosts, the imprint of the AGN-contamination channel, and every entry is significant at $p < 10^{-4}$. This opposes the sign in Figure~\ref{fig:drivers}, and the two are consistent because they condition on different quantities and isolate different channels: the figure measures host-only color at fixed stellar mass and finds over-massive systems redder, the stellar-population channel in which over-massive hosts are older and more quenched, while Table~\ref{tab:couplings} measures total color at fixed magnitude and finds over-massive systems bluer, the AGN-contamination channel in which a relatively brighter nucleus adds blue light. Because a predictor on conditionally-residualized inputs depends only on $P(\Delta \mid \delta_c, z)$, its per-domain conditioning medians are estimable from a survey without black hole labels. The host-only color coupling at fixed $M_*$ is present in \simba\ ($\rho = +0.118$ to $+0.289$ within the reference's stellar-mass support, Figure~\ref{fig:drivers}) and TNG100-1 ($+0.08$ to $+0.09$ at $z < 1.5$) but absent in \eaglesm\ ($|\rho| \leq 0.014$); in both simulations the coupling resides in the heavily attenuated tail at $g-r \geq 2$, with no consistent coupling in the blueward bulk, so the amplitude and the channel composition are simulation-dependent. Conditionally-residualized transfer collapses accordingly: \simba\ $\to$ TNG100-1 reaches $\rho = 0.048$ and \simba\ $\to$ \eaglesm\ reaches $\rho = 0.055$, so the invariant component is small and the raw baseline's $0.155$ rides on the non-invariant magnitude-redshift channel that does not survive into a different simulation. This universal sign is a prescription-independent prediction testable on sky data; its amplitude, the quantity a per-object predictor needs, is simulation-specific and would need to be calibrated against observations.

\begin{deluxetable}{lcccc}
\tabletypesize{\footnotesize}
\tablecaption{Conditional Color-$\Delta$ Couplings \label{tab:couplings}}
\tablehead{\colhead{Simulation} & \colhead{$\rho_{g-r}$} & \colhead{$\rho_{r-i}$} & \colhead{$\rho_{u-g}$} & \colhead{$N$}}
\startdata
\simba\ & $-0.051$ & $-0.058$ & $-0.030$ & $848{,}116$ \\
TNG100-1 & $-0.090$ & $-0.060$ & $-0.033$ & $15{,}242$ \\
\eaglesm\ & $-0.047$ & $-0.026$ & $-0.086$ & $25{,}123$ \\
\enddata
\tablecomments{Spearman rank correlation between the conditional color offset $\delta_c = c - {\rm median}(c \mid {\rm abs\text{-}mag\ decile} \times z\ {\rm bin})$ and $\Delta$, in cells of at least $100$ objects. All entries are significant at $p < 10^{-4}$. The negative sign is universal: over-massive black holes are bluer at fixed magnitude and redshift. The couplings are measured on the full per-simulation intrinsic frames cached for the Section~\ref{sec:cross_sim_results} transfer experiments before the $60{,}000$-row training subsample, so $N$ is the cached intrinsic sample size: a chunked $5$ percent draw of the \simba\ forward-model catalog and the full intrinsic TNG100-1 and \eaglesm\ catalogs. The cached \eaglesm\ frame carries 12 fewer rows ($0.05\%$) than the $N = 25{,}135$ primary-analysis sample of Table~\ref{tab:comparison}, a difference in row filtering between the two frames that is immaterial to the correlations. After the same conditional residualization, cross-simulation transfer falls to $\rho = 0.048$ (\simba\ $\to$ TNG100-1) and $\rho = 0.055$ (\simba\ $\to$ \eaglesm).}
\end{deluxetable}

\begin{table*}[tbp!]
\centering
\caption{Cross-Simulation Transfer, Intrinsic Mode \label{tab:cross_sim}}
\footnotesize
\setlength{\tabcolsep}{3pt}
\begin{tabular}{llcccc}
\toprule
\textbf{Direction} & \textbf{Config} & \textbf{$\sigma_{\rm res}$ Transfer (dex)} & \textbf{$\sigma_{\rm res}$ Within (dex)} & \textbf{$\sigma_{\rm pop}$ Target (dex)} & \textbf{Degradation (dex)} \\
\hline
\simba\ $\to$ TNG100-1 & Raw & 0.407 & 0.163 & 0.194 & $+0.244$ \\
\simba\ $\to$ TNG100-1 & Rank & 0.393 & 0.163 & 0.194 & $+0.230$ \\
TNG100-1 $\to$ \simba & Raw & 0.695 & 0.668 & 0.692 & $+0.027$ \\
TNG100-1 $\to$ \simba & Rank & 0.695 & 0.668 & 0.692 & $+0.027$ \\
\simba\ $\to$ \eaglesm & Raw & 0.521 & 0.171 & 0.180 & $+0.350$ \\
\simba\ $\to$ \eaglesm & Rank & 0.518 & 0.171 & 0.180 & $+0.347$ \\
\eaglesm\ $\to$ \simba & Raw & 0.690 & 0.668 & 0.692 & $+0.022$ \\
\eaglesm\ $\to$ \simba & Rank & 0.690 & 0.668 & 0.692 & $+0.022$ \\
TNG100-1 $\to$ \eaglesm & Raw & 0.188 & 0.171 & 0.180 & $+0.017$ \\
TNG100-1 $\to$ \eaglesm & Rank & 0.188 & 0.171 & 0.180 & $+0.017$ \\
\eaglesm\ $\to$ TNG100-1 & Raw & 0.296 & 0.163 & 0.194 & $+0.133$ \\
\eaglesm\ $\to$ TNG100-1 & Rank & 0.295 & 0.163 & 0.194 & $+0.132$ \\
\bottomrule
\end{tabular}
\tablecomments{Each row trains the XGBoost-plus-Random-Forest tree components on the source simulation and evaluates the residual scatter $\sigma_{\rm res}$ on the target. The within-simulation column is the same tree ensemble cross-validated on the target; the full stacked-ensemble within-simulation values are in Table~\ref{tab:comparison}. Degradation is the transfer minus within-simulation scatter and should be read relative to the target's $\sigma_{\rm pop}$. The small TNG100-1 $\to$ \eaglesm\ degradation of $0.017$ dex is an artifact of a near-constant prediction on the target (Pearson $r \approx -0.001$, Spearman $\rho \approx -0.004$); the raw cross-simulation Pearson $r$ is at most $0.155$ in every direction. The magnitude-limited mode follows the same pattern: degradation into \simba\ of $0.029$ to $0.043$ dex and out of \simba\ of $0.257$ to $0.350$ dex, with raw and rank again within $0.002$ dex of each other.}
\end{table*}

\section{Discussion} \label{sec:discussion}

\subsection{Interpretation}

The decomposition in Section~\ref{sec:circularity} is consistent with broadband colors carrying residual information about $\Delta$ at fixed $M_*$ predominantly through host stellar populations. The host SED encodes $\Delta$ at fixed $M_*$ through specific star formation rate and stellar age: \simba's residual correlates with sSFR \citep{Ma2025,thomas19}, and over-massive black holes formed their stars earlier than under-massive systems of the same mass \citep{martin18}, a signal the SPS model translates into broadband colors (Figure~\ref{fig:drivers}). The luminous high-redshift AGN that JWST finds in star-forming hosts \citep{Harikane2023,Maiolino2024,Pacucci2023} are selected on accretion, a population whose instantaneous specific star formation rate does not contradict the population-level trend toward redder, more quiescent hosts at fixed $M_*$ measured in this work.

Per-object $\Delta$ becomes useful once $\sigma_{\rm res}$ falls below the intrinsic scatter of the local scaling relations, roughly $0.3$ dex, where the photometric estimate becomes competitive with using the mean relation; \citet{KormendyHo2013} measure $0.29$ dex in $M_{\rm BH}$--$\sigma$. Broadband photometry reaches $\sigma_{\rm res} \approx 0.50$ to $0.65$ dex for \simba, so the gap to that threshold is a factor of $\sim 2$ in scatter, more than the measured color increments of $0.02$ to $0.03$ dex can close on their own. A single spectroscopic measurement of the stellar velocity dispersion closes most of that gap, because $M_{\rm BH}$--$\sigma$ predicts $M_{\rm BH}$ to $\sim 0.3$ dex and hence $\Delta$ at fixed $M_*$. Emission-line diagnostics that constrain specific star formation rate and stellar age sharpen the stellar-population channel that broadband colors only coarsely sample, so the spectroscopic and photometric channels are complementary.

\begin{figure}[tbp!]
\plotone{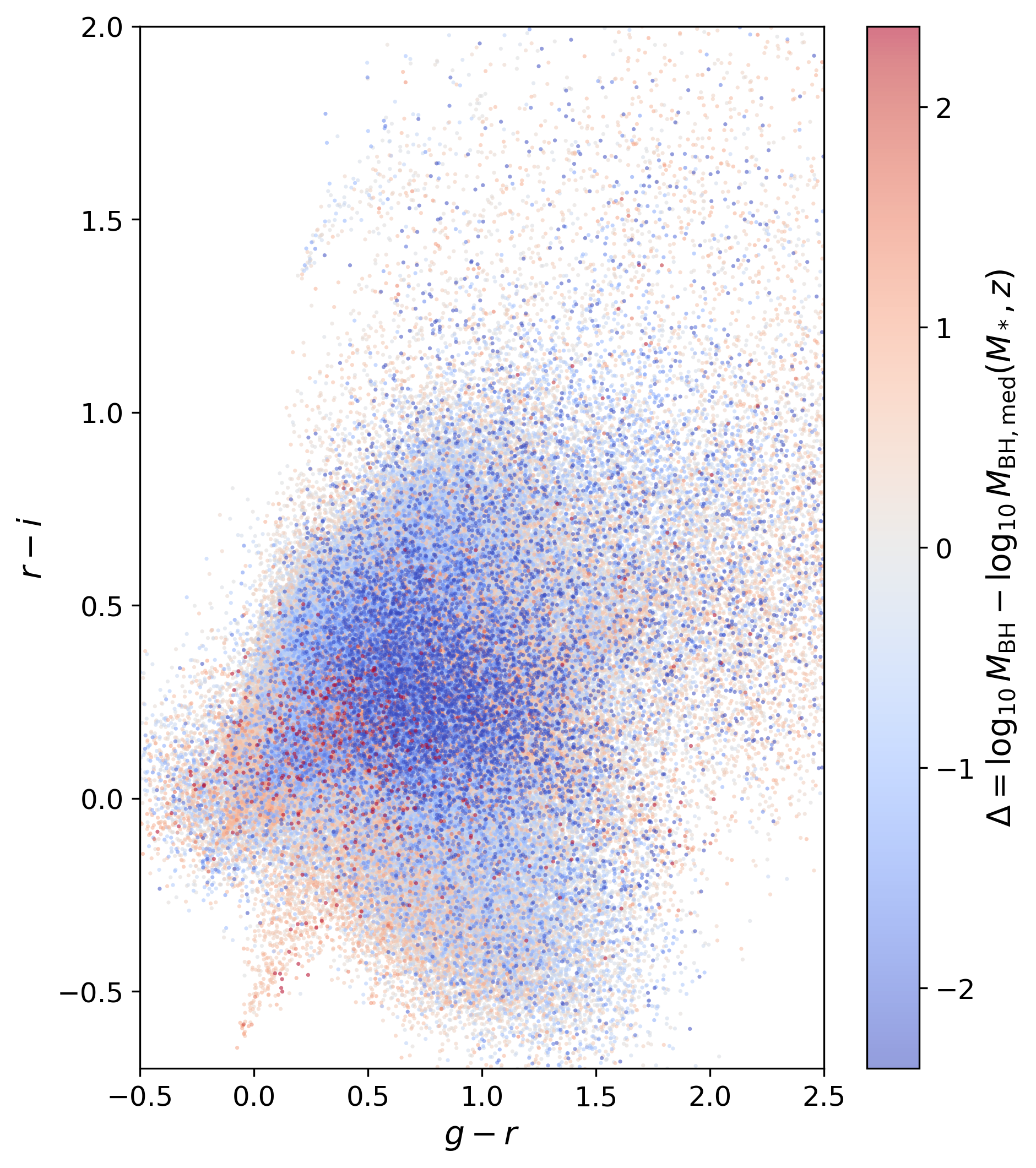}
\caption{Total $g-r$ versus $r-i$ color for a $200{,}000$-object \simba\ sample drawn from the forward-modeled catalog, with each source colored by its residual $\Delta$ on a diverging scale clipped at the 98th percentile of $|\Delta|$, and points drawn in order of ascending $|\Delta|$ so the strongest residuals sit on top. Over-massive (red) objects are displaced toward bluer colors in both $g-r$ and $r-i$, a separation driven by the tails, with median $(g-r, r-i)$ at $|\Delta| > 1$ of $(0.63, 0.23)$ for over-massive against $(0.82, 0.31)$ for under-massive objects, the separation the full-photometry regressor exploits. The direction matches the sign-universal negative fixed-magnitude color-$\Delta$ coupling of Table~\ref{tab:couplings}, which conditions on absolute-magnitude decile and redshift bin, whereas the plotted colors are unconditioned. Because $\Delta$ is negatively skewed (mean $-0.11$, median zero by construction), the symmetric scale saturates $\approx 1.8\%$ of points on the under-massive side against $\approx 0.2\%$ on the over-massive side, so the deep-blue extremes reflect this skew.}
\label{fig:color_overlay}
\end{figure}

The two conditional couplings are themselves the observational discriminator. At fixed magnitude and redshift the partial correlation between $\Delta$ and host color is negative and shared by all three simulations (Table~\ref{tab:couplings}; Figure~\ref{fig:color_overlay} shows the corresponding displacement in unconditioned colors), while at fixed stellar mass it is positive in \simba\ and \tng\ ($\rho \approx +0.08$ to $+0.29$; Figure~\ref{fig:drivers}) and absent in \eagle\ ($|\rho| \lesssim 0.014$); the positive coupling resides in the reddest hosts, so the fixed-mass test requires color coverage of the attenuated population. The routing of this coupling through the $g-r \geq 2$ tail is a prediction of the specific \citet{Calzetti2000} dust prescription (Section~\ref{sec:methods}); its red-tail fractions of $26.5\%$ and $66.7\%$ exceed those seen in observed galaxy color distributions, and alternative dust laws are among the forward-model systematics not treated in this work (Section~\ref{sec:limitations}). A sample with stellar masses from SED fitting and a $\Delta$ proxy, binned at fixed magnitude, should recover the negative fixed-magnitude coupling whichever simulation describes reality, while the positive-or-absent fixed-mass coupling is the second axis that separates the simulations. The standard error on a Spearman correlation is approximately $1/\sqrt{N-3}$, so a $3\sigma$ detection of the weakest universal coupling ($\rho \approx 0.05$) requires at least $N \approx 9/\rho^2$, a few thousand objects, while the stronger fixed-mass couplings ($\rho \approx 0.08$ to $0.29$) need $N \approx 110$ to $1{,}400$; this is a lower bound set by correlation statistics. A velocity-dispersion subset, in which $\sigma$ fixes $M_{\rm BH}$ to $\sim 0.3$ dex and hence $\Delta$ at fixed $M_*$ through the spectroscopic channel already invoked above, cross-matched to LSST imaging supplies such a sample, a few thousand objects from surveys like 4MOST, DESI, or SDSS-V.

These results have four implications for LSST-era analyses. A model claiming black hole recovery should benchmark against a regressor given the true stellar mass and redshift, without which stellar-mass recovery can be mistaken for black hole information. A model should not be trained on one simulation and deployed on sky data, since a \simba-trained model is anti-informative on TNG-like or \eagle-like populations and pooling did not recover transferable structure; only the sign-universal conditional color coupling can be calibrated against observed data. Inferences leaning on AGN brightness are simulation-dependent: the AGN channel's share of the population scatter varies by an order of magnitude between TNG100-1 and \simba\ (Table~\ref{tab:comparison}), and it differs threefold between the two Bondi-based simulations, so the coupling of nuclear light to $\Delta$ is set by more than the nominal mass dependence of the accretion prescription. Per-object $\Delta$ claims should be reserved for spectroscopic samples, imaging-only analyses targeting population-level statistics where the increments are resolved but small.

\subsection{Limitations and Future Directions} \label{sec:limitations}

Several limitations bound this forecast. First, the forward-modeled photometry is a re-encoding of catalog quantities ($M_*$, SFR, $\dot{M}$, $z$) through an SED recipe; channels that observed imaging captures beyond the recipe, including resolved morphology, variability, and spectral features outside the broadband bandpasses, are excluded by construction, so the forecast bounds the SED channel alone. LSST time-domain variability is a known additional AGN channel \citep{Ivezic2019} not represented in the forward model. Second, every regressor in the nested comparison receives the catalog redshift exactly; photometric-redshift error, which is the dominant uncertainty term for observed LSST AGN, would degrade every regressor and is not modeled. Third, no measurement-error scatter is injected into the recorded fluxes (Section~\ref{sec:methods}), so the measured increments are upper bounds with respect to photometric noise. Fourth, rescaling the per-band intrinsic-scatter amplitudes of the host SED model by factors of $0.5$ and $2.0$ moves the intrinsic full-photometry scatter from $0.6534$ to $0.6498$ and $0.6588$ dex and its increment over the host-magnitude proxy from $0.0321$ to $0.0350$ and $0.0273$ dex, and moves the magnitude-limited scatter from $0.5045$ to $0.4950$ and $0.5125$ dex and its increment from $0.0543$ to $0.0629$ and $0.0465$ dex; the increment keeps its sign and order of magnitude across this factor-of-four range in scatter amplitude. The intrinsic arm is a pure amplitude rescaling on the same underlying objects, whereas the magnitude-limited arm additionally folds in the survey-selection response, since sample membership itself shifts with the scatter amplitude, tying the measured color information to the assumed scatter amplitudes and to the forward-model systematics deferred below.

Because \simba, \tng, and \eagle\ employ heavy seed prescriptions, the recovery is within heavy-seed models; extending the framework to explicit light-seed channels (e.g., BRAHMA; \citealt{Bhowmick2025}) is needed to test early-universe seed-origin discrimination. Forward-model systematics (alternative SPS templates, dust attenuation laws, and mass-metallicity prescriptions) and realistic photometric-redshift uncertainties are not included in the present model.

\section{Conclusion} \label{sec:conclusion}

This work forecasts the information broadband LSST photometry carries about the residual $M_{\rm BH}$--$M_*$ offset $\Delta$ across cosmological hydrodynamic simulations. Our primary conclusions are:

\begin{itemize}

  \item We measure photometric recovery of $\Delta = \log_{10} M_{\rm BH} - \log_{10} M_{\rm BH,med}(M_*, z)$ on a nested set of regressors. For \simba\ at the intrinsic population level the residual scatter falls from $\sigma_{\rm pop} = 0.6966$ dex to $\sigma_{\rm res} = 0.6534$ dex with the full 27-feature photometry, a $6.2\%$ reduction in residual scatter, and in the magnitude-limited mode from $\sigma_{\rm pop} = 0.5982$ dex to $\sigma_{\rm res} = 0.5045$ dex, a $15.7\%$ reduction. The narrower-distribution \eagle\ and \tng\ volumes show the same recovery in both modes, with TNG100-1 the largest fractional reduction in the primary analysis; their per-volume scatters and channel splits are collected in Table~\ref{tab:comparison}.

  \item A regressor given the true $M_*$ and redshift recovers no residual mass information for \simba\ ($R^2 = -0.014$), confirming that the per-simulation per-redshift running-median construction orthogonalizes $\Delta$ against stellar mass (Section~\ref{sec:cross_val_prose}).

  \item The increment of the full six-band photometry over a single host flux for \simba\ intrinsic varies non-monotonically with redshift, dipping near $z \approx 2$--$3$ before rising toward $z > 5$, where the $M_{\rm BH}$--$M_*$ residual is largest (Section~\ref{sec:cross_val_prose}).

  \item A two-channel decomposition shows that host stellar populations dominate the photometric signal in the \simba\ intrinsic sample, with AGN light the subdominant channel; Figure~\ref{fig:drivers} traces this to the correlation between host $g-r$ color and $\Delta$ at fixed $M_*$, a coupling concentrated in the heavily attenuated red tail ($g-r \geq 2$). In the magnitude-limited sample the AGN channel grows as the survey retains AGN-bright objects (Section~\ref{sec:circularity}). For \simba\ the channel increments span $0.008$ dex (the intrinsic AGN channel) to $0.06$ dex (the magnitude-limited stellar-population channel over the no-recovery floor), well below the $\sim 0.3$ dex intrinsic scatter of the local scaling relations (Section~\ref{sec:discussion}).

  \item Broadband LSST photometry constrains population-level statistics of the $M_{\rm BH}$--$M_*$ residual but does not recover individual-galaxy $\Delta$ to better than a few tenths of a dex without spectroscopic follow-up. Within the adopted forward model, broadband six-band imaging measures what the SED channel alone conveys about $\Delta$; spectroscopic stellar age, sSFR, or velocity dispersion measurements will be needed to push beyond it.

  \item Cross-simulation transfer is partial and asymmetric, and its structure separates into a shared and a simulation-specific part. The sign of the conditional color-offset coupling is shared by all three simulations: at fixed magnitude and redshift, over-massive black holes sit in bluer hosts (Table~\ref{tab:couplings}), a simulation-agnostic prediction testable on sky data. Its amplitude is simulation-dependent and does not transfer across simulations: raw cross-simulation correlations are too weak for blind calibration to beat the target baseline, leave-one-out training does not recover transferable structure, and conditional residualization leaves only a small invariant coupling (Section~\ref{sec:cross_sim_results}). Simulation-trained per-object predictors therefore require calibration against observations.

\end{itemize}

\appendix

\section{\eagle\ Volume Statistics} \label{sec:appendix_eagle}

The RefL0025 results and both RefL0100 results are reported in Table~\ref{tab:comparison}; RefL0025 has the same numerical resolution as the flagship RefL0100 volume at a fraction of the compute cost, and the larger box supplies the statistics the small box cannot, both in the magnitude-limited mode ($N = 25{,}256$ against $2{,}649$) and in the intrinsic mode ($N = 254{,}199$ against $25{,}135$), turning the marginal RefL0025 increments into resolved positive ones in both modes.
\eagle's conservative Bondi-limited accretion produces lower median accretion rates than \tng, leaving few sources above the LSST co-added depths and a correspondingly small magnitude-limited sample ($N = 2{,}649$); the regressor given the true $M_*$ and redshift in that mode returns $R^2 = +0.017$, sharing the selection-shaped skewness signature of the flagship volume at lower amplitude (Section~\ref{sec:cross_val_prose}), with no residual stellar-mass information.

\section{Resolution-Artifact Runs (TNG100-2, TNG300-2)} \label{sec:appendix_resolution}

The lower-resolution TNG variants TNG100-2 and TNG300-2 yield the lowest $\sigma_{\rm res}$ values anywhere in this work, $0.140$ and $0.120$ dex in the intrinsic mode (Table~\ref{tab:resolution_artifact}), but these track a correspondingly narrow population scatter without recovering physical structure. Coarse black hole mass discretization at these resolutions compresses the intrinsic $\Delta$ distribution to $\sigma_{\rm pop} = 0.1808$ dex (TNG100-2) and $0.1721$ dex (TNG300-2); the seed-mass step sets the scale of the target, and the small absolute residuals follow from that compression.

The discretization leaves a second, more direct signature in the regressor given the true $M_*$ and redshift. For the TNG100-2 intrinsic mode this regressor returns $R^2 = +0.105$, above the band the running-median construction produces in every primary-analysis volume (the largest value elsewhere is $+0.031$, itself traced to selection-shaped skewness with no mass information; Section~\ref{sec:cross_val_prose}, Table~\ref{tab:comparison}). The same \code{GroupKFold} split by black hole identifier is used throughout, and the elevated value arises because $51.7\%$ of the TNG100-2 intrinsic black holes sit within $0.05$ dex of the seed mass, a degenerate sub-population that a regressor given $M_*$ and $z$ can partly recover, the same seed-pileup skewness at larger amplitude.

The TNG300-2 intrinsic increment makes the inflation explicit against a fixed comparison. Its full-over-single-flux increment is $0.0506 \pm 0.0006$ dex, roughly triple the $0.0174 \pm 0.0006$ dex measured for the primary-analysis TNG300-1 intrinsic mode (Table~\ref{tab:comparison}), and its own full-photometry fit reaches $R^2 = 0.508$, above every primary volume, the highest of which reaches $R^2 = 0.318$ (TNG100-1 intrinsic); both follow from the same discretized target that the regressor exploits in the absence of physical structure. The TNG300-2 magnitude-limited mode, with only $N = 2{,}649$ surviving the survey cut (coincidentally equal to the \eaglesm\ magnitude-limited count in Table~\ref{tab:comparison}), instead shows an AGN channel consistent with zero ($0.0011 \pm 0.0017$ dex), the expected behavior once the sample is too small to separate the channels. The resolution caveat that excludes both volumes from the primary analysis is therefore unchanged: coarse seed-mass discretization drives the $\Delta$ distribution.

\begin{table}[h]
\centering
\caption{Resolution-Limited TNG Runs \label{tab:resolution_artifact}}
\footnotesize
\setlength{\tabcolsep}{3pt}
\begin{tabular}{llrcc}
\toprule
\textbf{Simulation Volume} & \textbf{Mode} & \textbf{$N$} & \textbf{$\sigma_{\rm res}$ (dex)} & \textbf{$R^2$ (true $M_*$, $z$)} \\
\hline
TNG100-2 & Magnitude-Limited & 82,845 & $0.196$ & $+0.026$ \\
TNG100-2 & Intrinsic & 249,560 & $0.140$ & $+0.105$ \\
TNG300-2 & Magnitude-Limited & 2,649 & $0.222$ & $+0.029$ \\
TNG300-2 & Intrinsic & 245,837 & $0.120$ & $+0.023$ \\
\bottomrule
\end{tabular}
\tablecomments{Coarse black hole mass discretization at these resolutions produces a $\Delta$ distribution dominated by the seed-mass step; the small $\sigma_{\rm res}$ values reflect this narrow population scatter ($\sigma_{\rm pop} = 0.1808$ dex for TNG100-2 intrinsic, $0.1721$ dex for TNG300-2 intrinsic) without recovering physical structure. The TNG100-2 intrinsic value from the regressor given the true $M_*$ and redshift ($R^2 = +0.105$) sits above the near-zero band of the primary-analysis volumes because $51.7\%$ of its black holes lie within $0.05$ dex of the seed mass, an incomplete orthogonalization driven by the discretized target, with the same \code{GroupKFold} split by black hole identifier used throughout. These runs are excluded from the primary analysis.}
\end{table}

\begin{acknowledgments}
We thank the \simba\ team for making their simulation data publicly available. We gratefully acknowledge the IllustrisTNG collaboration for providing public access to simulation data via the TNG API and the EAGLE team for their public data releases.
The IllustrisTNG simulations were undertaken with compute time awarded by the Gauss Centre for Supercomputing (GCS) under GCS Large-Scale Projects GCS-ILLU and GCS-DWAR on the GCS share of the supercomputer Hazel Hen at the High Performance Computing Center Stuttgart (HLRS), as well as on the machines of the Max Planck Computing and Data Facility (MPCDF) in Garching, Germany.
We acknowledge the Virgo Consortium for making the EAGLE simulation data available. The EAGLE simulations were performed using the DiRAC-2 facility at Durham, managed by the ICC, and the PRACE facility Curie based in France at TGCC, CEA, Bruy\`{e}res-le-Ch\^{a}tel.
This work used data from LSST Data Preview~1 \citep{RubinDP1}, obtained during Rubin Observatory commissioning. This material is based upon work supported in part by the National Science Foundation through Cooperative Agreement AST-1258333 and Cooperative Support Agreement AST-1202910 managed by the Association of Universities for Research in Astronomy (AURA), and the Department of Energy under Contract No.\ DE-AC02-76SF00515 with the SLAC National Accelerator Laboratory managed by Stanford University.
We thank Ryan Farber for mentorship and guidance throughout this work.
\end{acknowledgments}

\section*{Data Availability}

The \simba\ simulation data are publicly available at \url{http://simba.roe.ac.uk/} \citep{Dave2019}. The IllustrisTNG simulation data are publicly available at \url{https://www.tng-project.org/data/} \citep{Nelson2019DataRelease}. The EAGLE simulation data are publicly available through the EAGLE database at \url{http://icc.dur.ac.uk/Eagle/database.php} \citep{McAlpine2016}. Rubin Observatory Data Preview~1 is available to data rights holders via the Rubin Science Platform \citep{RubinDP1}. The regression pipeline code and derived data products generated in this work will be shared on reasonable request to the corresponding author.

\vspace{5mm}
\facilities{Rubin (LSSTCam)}

\software{Python, NumPy \citep{harris2020}, SciPy \citep{virtanen2020}, Pandas \citep{mckinney2010}, Matplotlib \citep{hunter2007}, scikit-learn \citep{pedregosa2011}, PyTorch \citep{paszke2019}, XGBoost \citep{chen2016}, Astropy \citep{astropy2013,astropy2018,astropy2022}, h5py \citep{collette_h5py}, FSPS \citep{conroy09,conroy10}, \texttt{python-fsps} \citep{foreman_mackey_pythonfsps}}

\bibliography{main}{}
\bibliographystyle{aasjournalv7}

\end{document}